\documentclass[%
 reprint,
nofootinbib,
 amsmath,amssymb,
onecolumn,
]{revtex4-2}

\usepackage[normalem]{ulem}
\usepackage{amssymb}
\usepackage{amsthm}
\usepackage{amsmath}
\usepackage{graphicx}
\usepackage{dcolumn}
\usepackage{bm}
\usepackage{hyperref}
\usepackage[usenames,dvipsnames,x11names,table]{xcolor}
\usepackage{balance}
\usepackage{tikz}
\usetikzlibrary{shapes}
\usepackage{enumitem}
\setlist[itemize]{align=parleft,left=0pt..1em}
\providecommand{\keywords}[1]
{
  \small	
  \textbf{\textit{Keywords---}} #1
}

\renewcommand{\vec}[1]{\boldsymbol{\bm{#1}}}

\begin{document}


\title{A sharp interface approach for wetting dynamics of hydrophobic coated droplets and soft particles}

\author{Francesca Pelusi$^{1,\ddagger}$}
\email{f.pelusi@iac.cnr.it}
\author{Fabio Guglietta$^{1,\S}$}
\author{Marcello Sega$^2$}
\author{Othmane Aouane$^1$}
\author{Jens Harting$^{1,3}$}
\email{j.harting@fz-juelich.de}
\affiliation{$^1$~Helmholtz Institute Erlangen-Nürnberg for Renewable Energy (IEK-11), Forschungszentrum Jülich GmbH, Cauerstraße 1, 91058 Erlangen, Germany}
\affiliation{$^2$~Department of Chemical Engineering, University College London, London WC1E 7JE, United Kingdom}
\affiliation{$^3$~Department of Chemical and Biological Engineering and Department of Physics, Friedrich-Alexander-Universität Erlangen-Nüremberg, Cauerstraße 1, 91058 Erlangen, Germany}
\affiliation{$\ddagger$~Current affiliation: ~Istituto per le Applicazioni del Calcolo, CNR - Via dei Taurini 19, 00185 Rome, Italy}
\affiliation{$\S$~Current affiliation: ~Department of Physics \& INFN, Tor Vergata University of Rome, Via della Ricerca Scientifica 1, 00133, Rome, Italy}

\vspace{0.5cm}
\date{\today}

\begin{abstract}
The wetting dynamics of liquid particles, from coated droplets to soft capsules, holds significant technological interest. Motivated by the need to simulate liquid metal droplet with an oxidize surface layer, in this work we introduce a computational scheme that allows to simulate droplet dynamics with general surface properties and model different levels of interface stiffness, describing also cases that are intermediate between pure droplets and capsules. Our approach is based on a combination of the immersed boundary (IB) and the lattice Boltzmann (LB) methods. Here, we validate our approach against the theoretical predictions in the context of shear flow and static wetting properties and we show its effectiveness in accessing the wetting dynamics, exploring the ability of the scheme to address a broad phenomenology.
\end{abstract}

\maketitle

\section{\label{sec:intro}Introduction}

The wetting of a solid surface by a liquid coincides with its ability to preserve the contact with the liquid~\cite{DeGennes85,Brochard92,Bonn01,DeConinck01}. The wettability of a solid substrate by a pure droplet is quantified by the droplet's equilibrium contact angle $\theta_{eq}$, which, in turn, is determined by the balance between adhesive and cohesive forces of the three phases involved (solid, liquid, vapor). At the macroscopic scale, Young's equation~\cite{Young1805} describes this balance as
\begin{equation}\label{eq:young_eq}
\cos{\theta_{eq}} = \dfrac{\sigma_{sl}-\sigma_{sg}}{\sigma},
\end{equation}
where $\sigma_{sl}$, $\sigma_{sg}$ and $\sigma$ are the solid-liquid, solid-gas and liquid-gas surface , respectively. Eq.~\eqref{eq:young_eq} also estimates the degree of wettability, making the distinction between poor ($\theta_{eq} > 90^{\circ}$) and good ($\theta_{eq} < 90^{\circ}$) wetting regimes. Out of equilibrium, the additional complexities arising from time dependence and viscous dissipation make dynamic wetting critical to a wide range of phenomena including droplet spreading, capillary rise, imbibition, and more complex situations like fluid displacement in porous media or multiphase flow in oil recovery~\cite{andreotti2020statics,de2008wetting, bonn2009wetting, snoeijer2013moving}. The recent development of new catalytic devices, for example, requires the usage of liquid metals and metal alloys in the form of catalytic liquid droplets adsorbed on a porous solid support~\cite{Taccardi17,Hofer23}. However, several liquid metals such as gallium and gallium-based alloys oxidize when exposed to air and an inherent oxide layer appears on the top of the surface. This oxide layer acts as a solid-like ``skin'', encapsulating a liquid metal core~\cite{Daeneke18,Allioux22} and changes the wetting properties of the droplet~\cite{Jeyakumar11,Doudrick14,Joshipura21,zheng2022liquid}. Another example concerns the so called liquid marbles, realised by rolling a small liquid droplet in an poorly wetting powder. Because of the layer of powder grains at the liquid-air interface, the wetting of these droplets is inhibited~\cite{Bormashenko11,Ni2021}, as required in some recent technological and microfluidic applications~\cite{Zhao10,Avruamescu18}.
\begin{figure*}[t!]
    \centering
    \includegraphics[width=.95\linewidth]{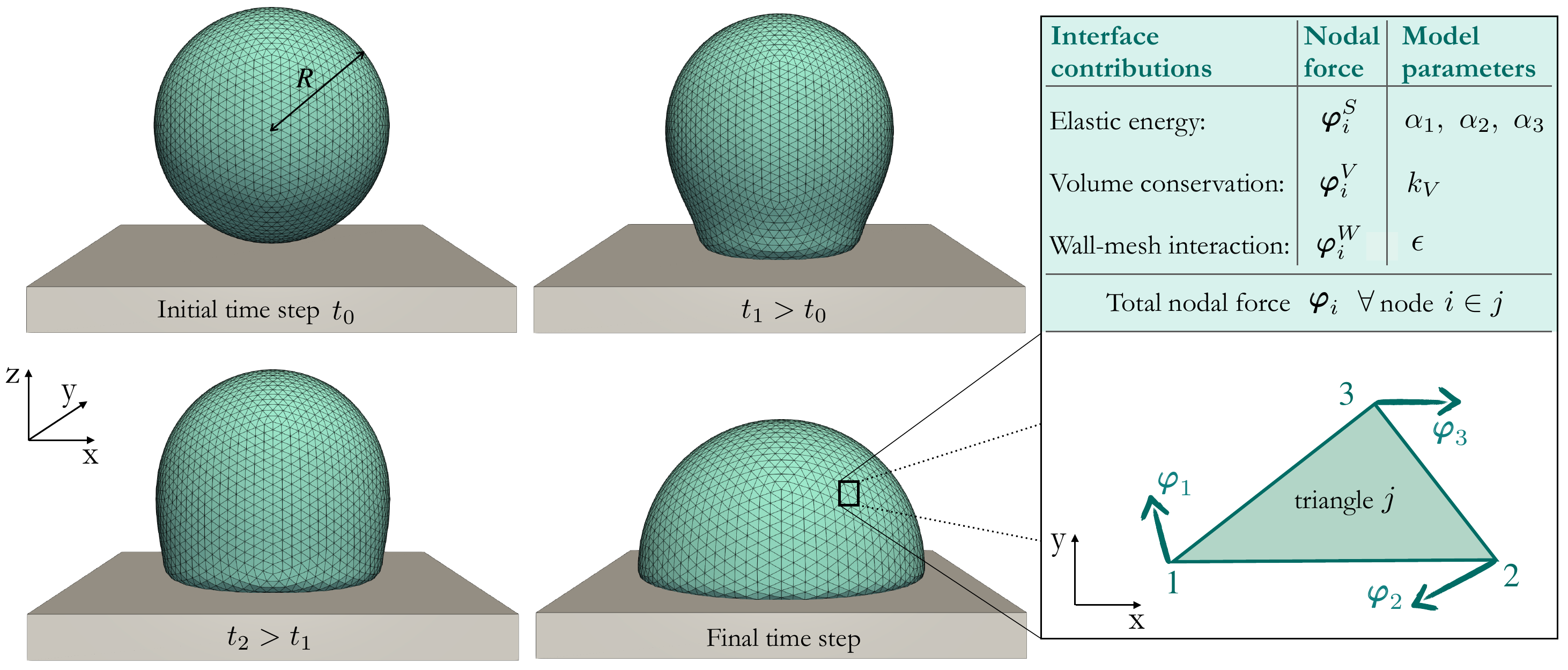}
    \caption{Sketch of the wetting dynamics of a generic particle with initial radius $R$ initially placed in contact with a flat wall. The interface is resolved with a 3D triangular mesh. On each triangular face $j$, some force contributions ${\pmb \varphi}$ are computed and distributed to the vertices $i \in j$, with the aim to consider (\textit{i}) the interface elasticity/rigidity, (\textit{ii}) the volume conservation, and (\textit{iii}) the wall-particle interaction. We also report the corresponding involved parameters.\label{fig:sketch}}
\end{figure*}
\begin{table*}[h!]
\begin{tabular}{| m{2.3cm} |m{3.2cm} | m{.6cm}| m{.6cm} |m{.5cm}| m{.5cm}|}
\hline
\multicolumn{6}{|c|}{Pre-stressed particles} \\
\hline
$\alpha = (\alpha_1,\alpha_2,\alpha_3)$ & Model  & $a$ & $b$ & $c$ & $d$\\
\hline 
\hline 
$\alpha = \Bar{\alpha} (1,0,0)$ & Pure droplet & 0.78 & 0.42 & 0.0 & 1.0\\
$\alpha = \Bar{\alpha} (1,0,1)$ & Softly coated droplet & 0.87 & 0.5 & 0.0 & 1.0\\
$\alpha = \Bar{\alpha} (1,1,1)$ & Rigidly coated droplet & 0.91 & 0.55 & 0.0 & 1.0\\
\hline
\end{tabular}
\hspace{0.08cm}
\begin{tabular}{| m{2.3cm} |m{3.5cm} | m{.55cm}| m{.5cm} |m{.5cm}| m{.55cm}|}
\hline
\multicolumn{6}{|c|}{Non-pre-stressed particles} \\
\hline
$\alpha = (\alpha_1,\alpha_2,\alpha_3)$ & Model  & $a$ & $b$ & $c$ & $d$\\
\hline 
\hline
$\alpha = \Bar{\alpha} (0,0,1)$ & Pure elastic capsule & 0.71 & 2.0 & 0.2 & 0.75\\ 
$\alpha = \Bar{\alpha} (0,1,1)$ &  Non-pre-stressed capsule & 0.7 & 1.6 & 0.2 & 0.75\\ 
& & & & &\\
\hline
\end{tabular}
\hspace{0.5cm}\caption{List of system models that can be explored by tuning the parameter $\alpha_1$, $\alpha_2$, and $\alpha_3$ in the interface model for a generic particle reported in Eq.~\eqref{eq:generalised_energy1}. The left and right tables refer to the pre-stressed and non-pre-stressed particle classes, respectively. For each model we display the corresponding fitting parameters $a$, $b$, $c$ and $d$ appearing in Eq.~\eqref{eq:cos_theta_prediction}.\label{table:cases}}
\end{table*}
The lattice Boltzmann (LB) method has been used since decades to address problems in wettability~\cite{Biferale07,Yan07,HuangSukop07,Attar09,Tanaka11,HKH11,Jansen13,Wang17,akai2018wetting,Zitz19,Wang20,Pelusi22}. A typical strategy used to simulate droplets within the framework of LB includes introducing non-ideal interface force models, such as the Shan-Chen~\cite{ShanChen93} and the Free-Energy ones~\cite{Swift96}. However, these approaches model a diffuse interface. Simulating droplets with a complex rheology, specifically coated droplets including for example liquid metal ones with an oxide layer or liquid marbles, require the use of a constitutive law for the interface. In this case it is more convenient to use a method that reproduces the sharp-interface limit of hydrodynamics. In addition, pseudopotential or free-energy approaches do not easily allow to model a behaviour that, as it is typical for coated droplets, is intermediate between the case of a pure droplet and that of a capsule, as is the case for coated droplets. For these reasons, here we have opted for combining the LB model with an immersed boundary (IB) method, which naturally preserves the hydrodynamic sharp-interface limit, to simulate the complex droplet's wetting dynamics (see Fig.~\ref{fig:sketch}).\\
Our goal is to introduce a comprehensive numerical approach that allows modelling droplets with complex interfacial properties in a consistent way. Here, we model the interface as a 3D triangular mesh and employ a constitutive law based on the theory of Barthès-Biesel and Rallison~\cite{barthes1981time} to explore the case of coated droplets. This approach allows us to describe in a continuous way the transition from pure droplet to capsule-like models by minimising the number of involved parameters.
We provide a validation of our IBLB numerical simulations against the theoretical prediction in case of a simple shear flow experiment. Then, we perform wetting dynamics simulations, which show a good agreement with experimental observations in the case of a pure droplet, and we explore the range of accessible contact angles in terms of the involved parameters and the intensity of the interaction with the wall. With this approach we aim at providing a qualitative approximation of the mechanical behaviour of droplets with a complete and wide range of interfacial properties. Nevertheless, the model can be further refined to include additional interface properties, enhancing its accuracy and applicability, such as an extended model to mimic the oxidized layer thickness when dealing with liquid metal droplets.

The paper is organised as follows: in Sec.~\ref{sec:membranemodel} we describe the  interface model introduced in the IBLB framework. Then, in Sec.~\ref{sec:method} we summarise the main features of the IBLB model employed. The benchmark of a droplet in a simple shear flow is shown in Sec.~\ref{sec:deformation}. In the context of the wetting dynamics, Sec.~\ref{sec:validation} report a model validation, while all wetting dynamics facets are analysed and discussed in Sec.~\ref{sec:results}. Results are summarised in Sec.~\ref{sec:conclusions}.

\section{Interface model\label{sec:membranemodel}}

In this section, we describe the theoretical model employed in this work to simulate a generic soft particle. Following Barthès-Biesel \& Rallison~\cite{barthes1981time}, we consider a two-dimensional, isotropic and homogeneous elastic interface with no bending resistance. Its mechanical response is characterised by an interface strain energy $w_S= w_S(I_1,I_2,\alpha_1,\alpha_2,\alpha_3)$ which is written in terms of the principal strain invariants $I_{1,2}$ and three parameters $\alpha_{1,2,3}$ as~\cite{barthes1981time}
\begin{widetext}
\begin{equation}\label{eq:generalised_energy1}
    w_S(I_1,I_2,\alpha_1,\alpha_2,\alpha_3)  = w_{S,0} +\frac{1}{2}(\alpha_1 - \alpha_3)\log(I_2+1)+\frac{1}{8}(\alpha_1+\alpha_2)\log^2(I_2+1) + \alpha_3 \left[ \dfrac{1}{2}(I_1+2)-1 \right],
\end{equation}
\end{widetext}
where $w_{S,0}$ is a reference value. $I_{1}$ and $I_{2}$ quantify the strain and dilation state of the membrane, respectively. The parameters $\alpha_{1,2,3}$, instead, characterize the material properties: The pre-stress $\alpha_1$ is an isotropic tension without an applied load; $\alpha_2$ is the resistance against area dilatation and $\alpha_3$ the resistance against shear deformation (i.e., the strain modulus). For the sake of simplicity, hereafter we will refer to these three parameters as $\alpha = (\alpha_1,\alpha_2,\alpha_3)$. Concerning their choice, we distinguish the two main classes of pre-stressed ($\alpha_1 > 0$) and non-pre-stressed ($\alpha_1 = 0$) particles. For the latter class, an appropriate combination of $\alpha_2$ and $\alpha_3$ leads to the well-known Skalak~\cite{skalakStrainEnergyFunction1973} and Neo-Hookean~\cite{matsunaga2016rheology} models. By conveniently tuning these parameters, one can switch from a pure droplet ($\alpha = (\sigma,0,0)$, where $\sigma$ is the surface tension) to a pure elastic capsule ($\alpha = (0,0,\alpha_3 > 0)$) and describe intermediate and more complex situations. In particular, in this work we consider also the classes of particles with $\alpha= (\alpha_1 >0, 0, \alpha_3 >0)$ and $\alpha=(\alpha_1 >0, \alpha_2 >0, \alpha_3>0)$. We call the first type of particle a ``softly coated droplet'' because it has the characteristic surface tension term of a droplet, but also a strain modulus because of $\alpha_3>0$. Because of the presence of a dilatational term $\alpha_2>0$, we call the second type ``rigidly coated droplet''. Here we argue that the latter case could be used to describe the interfacial properties of particles like liquid metal droplets with  oxidized surface. In Table~\ref{table:cases} we summarise the type of particles investigated in this work.\\
In order to check that Eq.~\eqref{eq:generalised_energy1} leads to the correct particle dynamics, we consider the case of a shear flow experiment, where a particle with initial radius $R$ and dynamic viscosity $\mu$ is placed between two distant moving walls which generate a shear rate $\dot{\gamma}$ (see top panel of Fig.~\ref{fig:deformation}). In this setup, time-dependent motion of the particle shape may be distinguished into two contributions: a solid body rotation and a stretching. Notice that the interface may rotate via a tank-treading motion despite of the particle reaching its steady state. This means that at this stage the interface deformation is constant in time at an Eulerian point $\vec{x}$, but it is not constant by looking at a point $X$ on the interface. Thus, following concepts and notation of Ref.~\cite{barthes1981time}, the time-evolution of the dimensionless position $\vec{x}$ (i.e., the position divided by the initial radius $R$), which is representative of the deformation field, reads
\begin{equation}\label{eq:BB81_1}
\vec{x} = \vec{X} + \beta \left[ \vec{K} \cdot \vec{X} + \vec{XX}\cdot \left(\vec{J}-\vec{K}\right)\cdot \vec{X} \right],
\end{equation}
where $\beta \ll 1$ is the expansion coefficient around the initial spherical position (we truncate the equation at the leading order in $\beta$), while $\vec{J}$ and $\vec{K}$ are two symmetric and traceless second-rank tensors which depend only on time. It follows that the instantaneous external shape of the
particle $r$ can be computed in terms of the norm of Eq.~\eqref{eq:BB81_1}, together with its normal $\vec{n}$~\cite{barthes1981time},
\begin{subequations}\label{eq:BB81_2}
\begin{gather}
    r \equiv |\vec{x}| = 1 + \beta \vec{X} \cdot \vec{J} \cdot \vec{X} = 1 + \beta \dfrac{\vec{x} \cdot \vec{J} \cdot \vec{x}}{r^2},\\
    \vec{n} = \dfrac{\vec{x}}{r} + 2 \beta \left[ \dfrac{\vec{xx}\cdot \vec{J} \cdot \vec{x}}{r^3} - \dfrac{\vec{J}\cdot \vec{x}}{r}\right] \ .
\end{gather}
\end{subequations}
Since in Eqs.~\eqref{eq:BB81_2} only tensor $\vec{J}$ appears, it means that 
$\vec{J}$ describes the overall deformation (i.e., the stretching contribution), while $\vec{K}$ describes the motion on the interface (i.e., the solid body rotation)~\cite{barthes1981time}. We remark that $\vec{J}$ is traceless because of the volume conservation constraint, whereas this property for $\vec{K}$ can be checked \textit{a posteriori}. In the limit of small deformations, the evolution equations for $\vec{J}$ and $\vec{K}$ are given by~\cite{barthes1981time}
\begin{equation}\label{eq:evJK}
\begin{cases}
\displaystyle \frac{\mathfrak{D}}{\mathfrak{D}t} \vec{K}  = \frac{5 \vec{E}}{2 \lambda + 3} + \frac{\vec{L}}{2 \lambda + 3} + \frac{\vec{M} \left(6 \lambda + 4\right)}{\left(2 \lambda + 3\right) \left(19 \lambda + 16\right)}\\
\displaystyle \frac{\mathfrak{D}}{\mathfrak{D}t} (\vec{J}-\vec{K}) =\frac{2}{19\lambda+16}\vec{M},
\end{cases}
\end{equation}
where  
\begin{equation}
    \frac{\mathfrak{D}\vec{A}}{\mathfrak{D}t} = \frac{d\vec{A}}{dt} 
- ({\bm \Omega} \cdot \vec{A} - \vec{A} \cdot {\bm \Omega})
\end{equation}
is the Jaumann derivative\cite{barthes1981time} applied to a generic tensor $\vec{A}$, which takes into account the rotation of the particle with the vorticity of the external fluid. $\vec{E}$ and ${\bm \Omega}$ are the symmetric and asymmetric part of the velocity gradient, respectively, $\lambda$ is the viscosity ratio between inside and outside fluids, and
\begin{subequations}\label{eq:L_M}
\begin{gather}
\vec{L} = 4(\alpha_2+\alpha_3)\vec{J} - (6\alpha_2 + 10\alpha_3)\vec{K}, \\
\vec{M} = -4(\alpha_1+2\alpha_2+2\alpha_3)\vec{J} + (12\alpha_2 + 16\alpha_3)\vec{K} \ .
\end{gather}
\end{subequations}
By numerically integrating Eq.~\eqref{eq:evJK}, it is possible to obtain information on the transient deformation dynamics since the tensor $\vec{J}$ is directly related to the particle-deformation as~\cite{BarthesSgaier85}
\begin{equation}\label{eq:deformation_J}
D = \mbox{Ca} (J_{11}^2+J_{12}^2)^{1/2},    
\end{equation}
where Ca$= \mu R \dot{\gamma} / \alpha$ is the capillary number. In the latter definition, one can consider $\alpha=\alpha_1=\sigma$ for a pure droplet or $\alpha=\alpha_3$ for a pure capsule. For the sake of simplicity, we fix the values of parameters $\alpha_1$, $\alpha_2$, and $\alpha_3$ to be equal to the same value $\Bar{\alpha}$ and we will refer to this triad of values simply as $\alpha = \Bar{\alpha} (0/1,0/1,0/1)$, with the vector elements turned on (1) and off (0) with the corresponding model (see Table~\ref{table:cases}).


\section{\label{sec:method}Numerical implementation}

The dynamics of the inner and outer fluid is simulated using a single-component lattice Boltzmann (LB) method in terms of the fluid particle populations $f_i(\hat{{\bf x}},t)$. The latter represents the probability distribution function of finding a fluid particle in a discrete lattice (Eulerian) node $\hat{{\bf x}}$ at a discrete time $t$. The corresponding macroscopic behaviour is recovered in the long-wavelength limit, which allows the link with the Navier-Stokes equations. Indeed, the solutions of the Navier-Stokes equation for the total density and momentum are easily accessible from the populations as $\rho(\hat{{\bf x}},t) = \sum_i f_i(\hat{{\bf x}},t)$ and  $\rho(\hat{{\bf x}},t){\bm u}(\hat{{\bf x}},t) = \sum_i {\bm c}_i f_i(\hat{{\bf x}},t)$, respectively, with ${\bm c}_i$ representing a set of 19 discrete velocities ($i = 0, \dots, 18$) living on a three-dimensional lattice (i.e., we employ a D3Q19 LB model). The dynamics of $f_i$ is ruled by a continuous succession of propagation and collision steps, as highlighted by the discretized Boltzmann equation~\cite{benzi1992lattice,Kruger16}
\begin{widetext}
\begin{equation}\label{eq:LBeq}
f_i(\hat{{\bf x}}+{\bm c}_i  \Delta t, t+\Delta t) - f_i(\hat{{\bf x}},t)  = - \dfrac{\Delta t}{\tau}\left[ f_i(\hat{{\bf x}},t) -f_{i}^{(eq)}(\hat{{\bf x}},t)\right] + w_i \left( 1 - \frac{\Delta t}{2 \tau}\right) \left( \frac{(\vec{c}_i-\vec{u})\cdot\vec{\mbox{F}}}{c_s^2} + \frac{(\vec{c}_i\cdot\vec{\mbox{F}})(\vec{c}_i\cdot\vec{u})}{c_s^4}\right) \Delta t,
\end{equation}
\end{widetext}
where $\Delta t$ is the time step. The propagation of $f_i$ on the lattice is described by the l.h.s. of Eq.~\eqref{eq:LBeq} with the help of ${\bm c}_i$, while the single-relaxation-time BGK approximation of the collision operator appears as the first term in the r.h.s. The latter has the aim of modelling the relaxation of $f_i$ towards the equilibrium distribution $f_{i}^{(eq)}(\hat{{\bf x}},t)$, represented as the local Maxwellian distribution
\begin{equation}\label{eq:feq}
\small
f_{i}^{(eq)}(\hat{{\bf x}},t) = w_i \rho \left[1+\frac{u_{k} c_{i,k}}{c^2_s} + \frac{u_{k} u_{j} (c_{i,k} c_{i,j} - c^2_s \delta_{kj})}{2 c^2_s} \right].
\end{equation}
The relaxation process lasts for a relaxation time $\tau$. In Eq.~\eqref{eq:feq}, $f_{i}^{(eq)}$ is weighted by the lattice-dependent weights $w_i$~\footnote{The weight $w_i$ in the employed D3Q19 model are $w_i = 1/3$ for $i =0$, $w_i = 1/18$ for $i =1\ldots6$, $w_i = 1/36$ for $i =7\ldots18$.} and depends on the speed of sound $c_s=\Delta \hat{\mbox{x}}/(\sqrt{3}\Delta t)$, where $\Delta \hat{\mbox{x}}$ is the lattice spacing. The last term of Eq.~\eqref{eq:LBeq} refers to the forcing implementation following the Guo scheme~\cite{Guo02}, where ${\bf F}$ is the force acting on the fluid. Notice that, to guarantee the second-order space-time accuracy, this forcing scheme modifies the fluid velocity as $\rho(\hat{{\bf x}},t){\bm u}(\hat{{\bf x}},t) = \sum_i {\bm c}_i f_i(\hat{{\bf x}},t) + {\bf F}\Delta t /2$. In our simulations, we keep fixed to unity both $\Delta \hat{\mbox{x}}$ and $\Delta t$. Furthermore, the fluid dynamic viscosity $\mu$ in LB models is related to the relaxation time $\tau$ as $\mu = c_s^2 \rho (\tau - 1/2)$. Here, we keep the viscosity ratio $\lambda$ fixed to unity, since the investigation on the role played by $\lambda$ goes beyond the purpose of this work. 

Then, to simulate the interface of a coated droplet or soft particle immersed in the surrounding LB fluid, we model the spherical particle interface using a 3D triangular mesh generated from a recursive refining of an icosahedron. Thus, the mesh resolution is defined in terms of the total number of triangular faces $N_{f}$ (see Fig.~\ref{fig:resolution_test} for a pictorial view of particles with different resolutions). To couple the soft particle dynamics with that of the surrounding fluid, we use the immersed boundary (IB) method, i.e., a fluid-mesh interaction method developed for the first time by Peskin~\cite{Peskin02} and based on the distinction between interface (Lagrangian) nodes ${\bf q}(t)$ and fluid (Eulerian) nodes $\hat{{\bf x}}$. The resulting coupling is distinct in two operations, i.e., interpolation and spreading. The interpolation operation consists of the computation of the $i-$th interface-node velocity ${\bf \dot{q}}_i(t)$ from the fluid one (${\bf u}(\hat{{\bf x}},t)$) as~\footnote{Note that Eq.~\eqref{eq:ibm-governing_eq1} causes the velocity of the surface to be equal to the fluid velocity, ensuring thus the no-slip boundary condition at the interface~\cite{Kaoui11,Kruger16}.}~\cite{Kruger16}
\begin{equation}\label{eq:ibm-governing_eq1}
\dot{{\bf q}}_i(t) = \sum_{\hat{{\bf x}}} {\bf u}(\hat{{\bf x}},t) \delta_D (\hat{{\bf x}}-{\bf q}_i(t)) \Delta \hat{\mathrm{x}}^3.
\end{equation}
This operation allows to update the node position ${\bf q}_i(t)$ as:
\begin{equation}\label{eq:ibm_update_pos}
    {\bf q}_i(t+\Delta t) = {\bf q}_i(t) + \dot{{\bf q}}_i(t)\Delta t.
\end{equation}
Then, the spreading operation is an interpolation of the interface nodal force to the fluid one which allows to make the latter aware of the presence of the interface: at this step, the total force (volume-)density the particle exerts on the fluid at the Eulerian node $\hat{{\bf x}}$ is given by
\begin{equation}\label{eq:ibm-governing_eq2}
    {\bf F}(\hat{{\bf x}},t) = \sum_i {\pmb{\varphi}}_i(t)\delta_D(\hat{{\bf x}}-{\bf q}_i(t)),
\end{equation}
where ${\pmb \varphi}_i$ is the total force on the Lagrangian node $i$ and the sum runs over all Lagrangian nodes. Both operations involve the so-called discrete delta function $\delta_D$, which is used to approximate the Dirac delta function on our lattice and is defined as~\cite{Peskin02,Kaoui11,Kruger16} 
\begin{equation}
    \delta_D(\hat{{\bf x}}) = \frac{1}{\Delta \hat{x}^3}\phi_4(\hat{\mathrm{x}})\phi_4(\hat{\mathrm{y}})\phi_4(\hat{\mathrm{z}})\ ,
\end{equation}
where $\phi_4(\mathrm{r})$ is the ``interpolation stencil'' involving four Eulerian nodes along each coordinate axis~\cite{kruger2011efficient} and defined as follows:
\begin{equation}\label{eq:ibm-interpolation_stencil}
\small
\phi_4(\hat{\mathrm{x}}) = \begin{cases}
\frac{1}{8} \left(3 - 2 \vert \hat{\mathrm{x}}\vert + \sqrt{1+4\vert \hat{\mathrm{x}}\vert-4\hat{\mathrm{x}}^2}\right)& \hspace{0.2cm} 0\le \vert \hat{\mathrm{x}}\vert \\
\frac{1}{8} \left(5 - 2 \vert \hat{\mathrm{x}}\vert - \sqrt{-7+12\vert \hat{\mathrm{x}}\vert-4x^2}\right)
 & \hspace{0.2cm} \Delta \hat{\mathrm{x}} \le \vert \hat{\mathrm{x}}\vert \le 2\Delta \hat{x} \\
 0 & \hspace{0.2cm} 2\Delta \hat{\mathrm{x}} \le \vert \hat{\mathrm{x}}\vert 
\end{cases}
\end{equation}
The resulting IBLB method has been largely used to simulate the dynamics of capsules~\cite{kruger2011efficient,aouaneStructureRheologySuspensions2021,BAHK21,KKH14,guglietta2023suspensions} and red blood cells~\cite{krugerDeformabilitybasedRedBlood2014,kaouiHowDoesConfinement2012,liSimilarDistinctRoles2021,guglietta2020lattice,guglietta2021loading}. However, only a few works employed this method also for simulating droplet dynamics~\cite{liFiniteDifferenceMethod2019,Guglietta2020,taglientiReducedModelDroplet2023}. A detailed step-by-step description of the IBLB algorithm implementation can be found in Ref.~\cite{Kruger16}.

In our implementation, the total nodal force ${\pmb \varphi}_i$, appearing in Eq.~\eqref{eq:ibm-governing_eq2} and acting on the $i$-th node at position $\vec{\mathrm{r}}_i$ at time $t$, is given by the sum of several contributions, i.e.,
\begin{equation}\label{eq:total_nodal_force}
    {\pmb \varphi}_i = {\pmb \varphi}^{S}_i + {\pmb \varphi}^{V}_i + {\pmb \varphi}^{\mbox{\tiny W}}_i.
\end{equation}
Each contribution plays a distinct role. First of all, ${\pmb \varphi}^{S}_i$ incorporates the information on the elastic properties of the interface. Thus, we compute this nodal force term as
\begin{equation}\label{eq:nodal_force_S}
    {\pmb \varphi}^S_i = - \dfrac{\partial}{\partial \{ {\bf q}_i \}}w_S(\{ {\bf q}_i \}),
\end{equation}
where $w_S$ is the generalised strain energy defined in Eq.~\eqref{eq:generalised_energy1}. Eq.~\eqref{eq:nodal_force_S} is calculated using a first order finite element method as described in Ref. ~\cite{kruger2011efficient}. Then, because we are dealing with incompressible fluids, we need to consider a volume conservation constraint. With this aim, we follow Ref.~\cite{kruger2011efficient} and we write the nodal volume force contribution ${\pmb \varphi}^V_i$ as~\cite{KrugerPhDthesis12}
\begin{equation}\label{eq:nodal_force_V}
    {\pmb \varphi}^V_i = - \dfrac{\partial }{\partial \{ {\bf q}_i \}}w_V(\{ {\bf q}_i \}),
\end{equation}
where $w_V = k_V(V-V_0)^2/2V_0$ is the volume energy. In this definition of the volume energy, $k_V$ refers to the volume-force coefficient and it is kept fixed to $1$. $V = \sum_j V_j$ is the instantaneous total particle volume, with the index $j$ running over the number of faces $N_f$~\footnote{Note that $V$ is functionally dependent on $\{ {\bf r}_i\}$.}, while $V_0$ is the initial total particle volume. Further details on how to compute nodal force contributions in Eqs.~\eqref{eq:nodal_force_S} and~\eqref{eq:nodal_force_V} can be found in Ref.~\cite{KrugerPhDthesis12}. The last contribution in Eq.~\eqref{eq:total_nodal_force} corresponds to the wall-particle interaction, the key element for wetting dynamics simulations. 
The IBLB approach used in this work involves only one single fluid component and it is not possible to control the wall-fluid surface tensions $\sigma_{sl}$ and $\sigma_{sg}$ by introducing two different interactions as, for example, Huang and coworkers did in the case of multi-component pseudopotential LB models~\cite{HuangSukop07}. However, many implementation of fluid-wall interactions in the case of single-component LB models~\cite{martys1996,HKH06,li2014contact} use a pseudopotential-like fluid-wall interaction which does not control $\sigma_{sl}$ and $\sigma_{sg}$ separately, but only their overall effect. These models work well in describing the wetting dynamics of droplets, despite of this limitation. In this work, we follow the same approach and we introduce a Lennard-Jones interaction on behalf of the wall-particle interaction:
\begin{equation}\label{eq:LJ_interaction}
    {\pmb \varphi}^{\mbox{\tiny W}}_i = 48\epsilon \left[\left(\dfrac{\xi}{d_i}\right)^{12}-\dfrac{1}{2}\left(\dfrac{\xi}{d_i}\right)^6\right]\dfrac{{\bf d}_i}{d_i^2},
\end{equation}
where ${\bf d}_i$ is the shortest displacement vector between the centroid of the triangle to which node $i$ belongs and the wall surface, and $d_i = |{\bf d}_i|$. It means that the force is computed once for each triangle, and then it is distributed on its vertices.  The choice of employing a Lennard-Jones interaction potential results from the necessity to model adhesive and repulsive forces between the droplet or particle and the surface. It follows the large amount of works based on molecular dynamics simulations which successfully studied the behaviour of nanodroplets or ridges on chemically patterned substrates~\cite{koplik1995,koplik2000,koplik2006,Semiromi11,DRKHD12}. Notice that in this work, we set $\xi = 0.5\Delta \hat{\mathrm{x}}$ to have an interface-wall interaction that decays to zero after one lattice spacing, thus respecting as much as possible the microscopic range, and tune the interaction by changing only $\epsilon$. The nodal force contributions along with the corresponding parameters are summarised in Fig.~\ref{fig:sketch}.

By summarizing, the IBLB algorithm implemented in this work matches the following steps~\cite{Kruger16}:
\begin{enumerate}[itemsep=0em,leftmargin=*]
    \item Compute the nodal force ${\bm \varphi}_i$ on each node $i$ (Eq.~\eqref{eq:total_nodal_force});
    \item Spread the nodal force to obtain the force acting on the fluid ${\bf F}(\hat{{\bf x}},t)$ via Eq.~\eqref{eq:ibm-governing_eq2};
    \item Perform the LB integration step: compute equilibrium distributions (Eq.~\eqref{eq:feq}), then apply the collision and perform the propagation. At this stage, ${\bf F}(\hat{{\bf x}},t)$ enters in r.h.s. of Eq.~\eqref{eq:LBeq};
    \item Compute the fluid velocity ${\bm u}(\hat{{\bf x}},t)$ from LB populations;
    \item Interpolate the fluid velocity to compute the Lagrangian node velocity (Eq.~\eqref{eq:ibm-governing_eq1});
    \item Update the position of each node ${\bf q}(t)$ via Eq.~\eqref{eq:ibm_update_pos};
    \item Iterate from step 1.
\end{enumerate}
All simulations have been performed in a periodic domain in the x- and y- directions, while two walls are placed along the (vertical) z-direction. A half-node bounce-back rule implements second-order no-slip boundary conditions at the walls~\cite{Kruger16}. Dimensional quantities are shown in lattice Boltzmann units (lbu).

Note that, although this method is not able to capture particle breakup and coalescence, it provides an easy way to model different systems by simply tuning the $\alpha_i$ parameters, as detailed in Section~\ref{sec:membranemodel}. The introduction of the wall-particle interaction~\eqref{eq:LJ_interaction} induces an accumulation of interface nodes on the wall-particle contact area. Such an aggregation is more prominent for high values of $\epsilon$ that are required for observing small contact angles, and is responsible for a numerical instability in that regime. Re-meshing technique may help mitigating this problem, but since we are interested in large contact angles, this accumulation does not affect the results presented in this paper.
\section{Benchmark: shear flow dynamics}\label{sec:deformation}
\begin{figure*}[t!]
    \centering
    \begin{tabular}{c}
         \includegraphics[width=.75\linewidth]{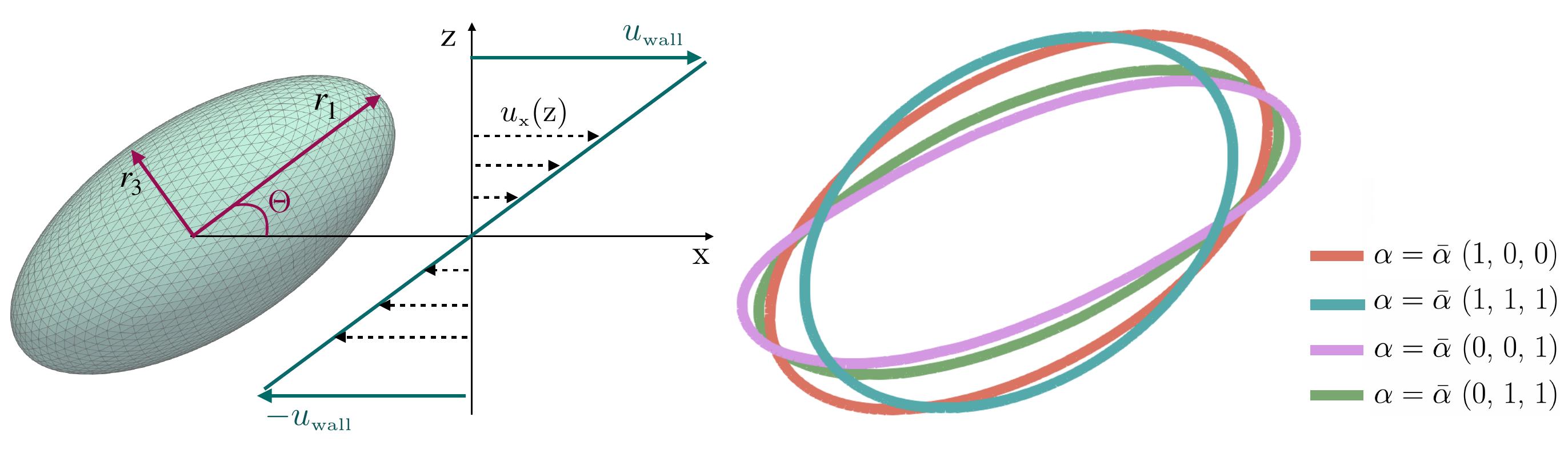}\\
         \includegraphics[width=.95\linewidth]{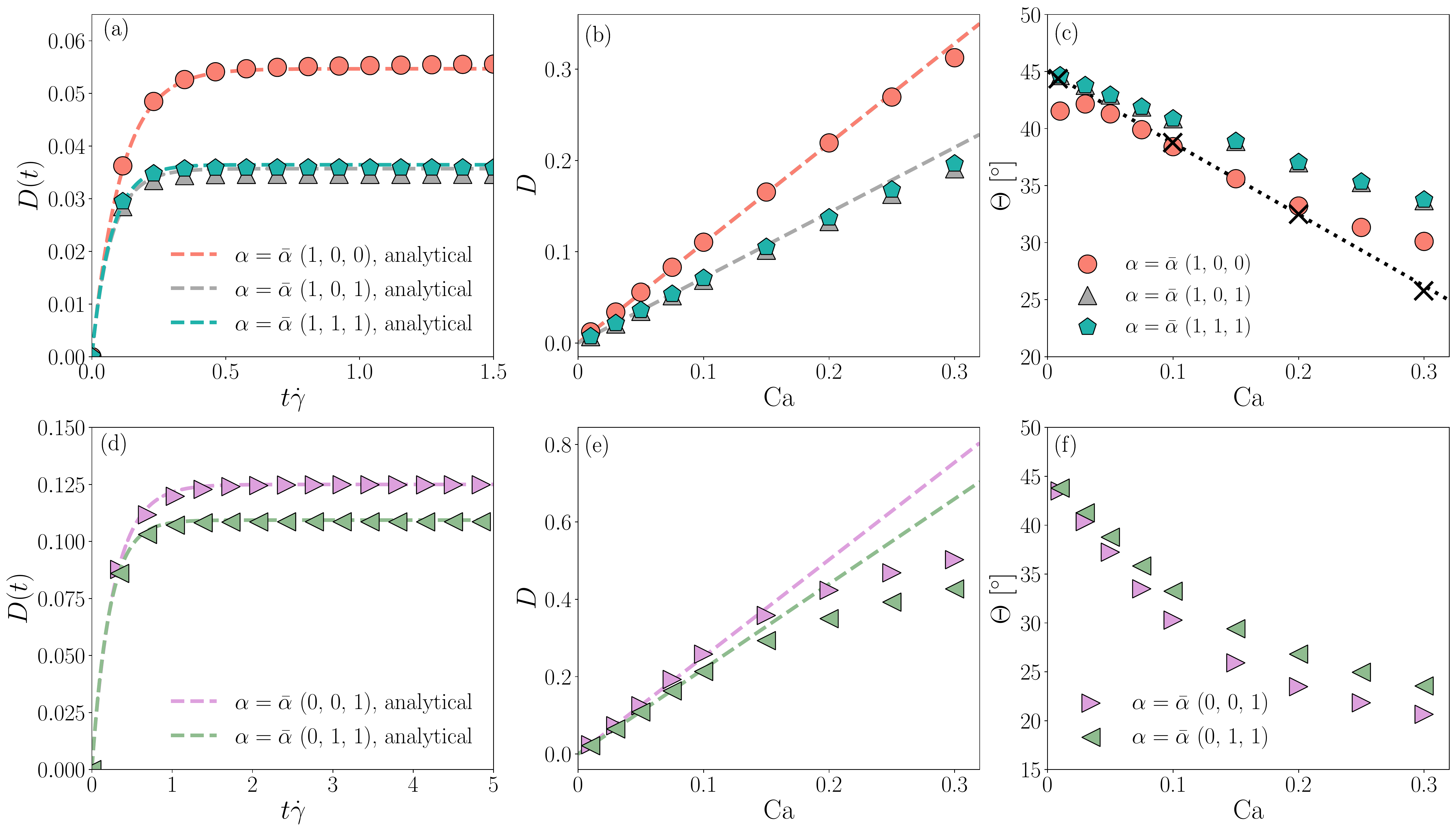} \\
    \end{tabular}
    \caption{Simulated experiment of a single particle under shear flow for the pre-stressed (panels (a)-(c)) and non-pre-stressed (panels (d)-(f)) particle models. In all panels, different symbols/colours refer to different models.  Top panels: a sketch of the shear experiment and final shape of the particle for Ca$=0.3$. Panels (a) and (d): time evolution of the deformation index $D(t)$ (Eq.~\eqref{eq:taylor_deform}) as a function of time for capillary number Ca=0.05. Time is shown normalised with the shear rate $\dot{\gamma}$, and dashed lines refer to the analytical solutions of Eqs.~\eqref{eq:evJK}. Panels (b) and (e): the steady-state value of the deformation $D$ as a function of Ca. Dashed lines draw the theroretical predictions: Eq.~\eqref{eq:deformation_droplet} (salmon line) for the pure droplet case and Eq.~\eqref{eq:deformation}  (other colors lines) for all the other models. Panels (c) and (f): the steady-state value of the inclination angle $\Theta$ as a function of Ca. To validate the model in the case of a pure droplet, we report black crosses from Ref.~\cite{Gounley16}, and we draw dotted lines for Eq.~\eqref{eq:inclination_angle}.\label{fig:deformation}}
\end{figure*}
\begin{figure*}[t!]
    \centering
     \begin{tabular}{c c}
    \includegraphics[width=.45\linewidth]{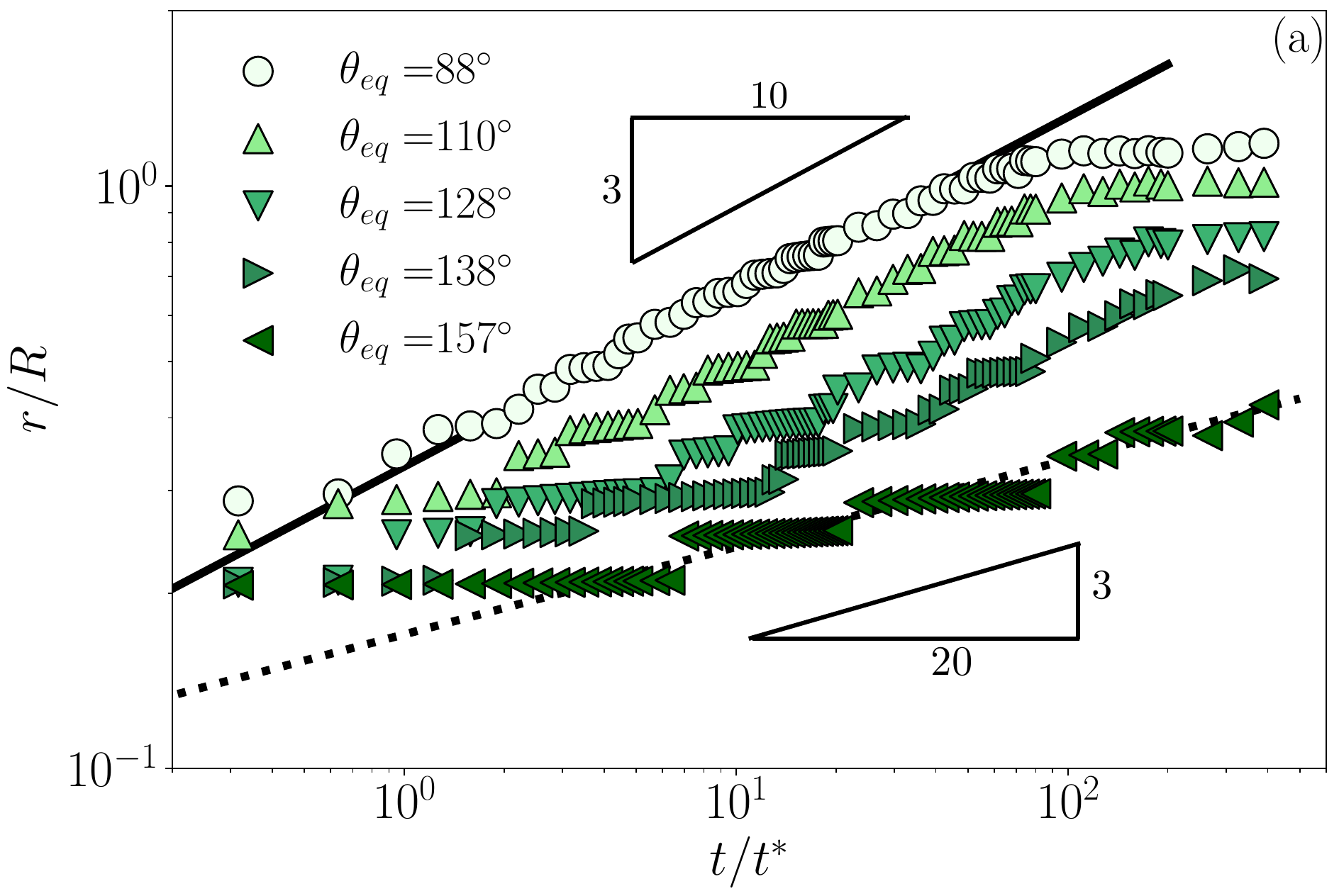} & \includegraphics[width=.453\linewidth]{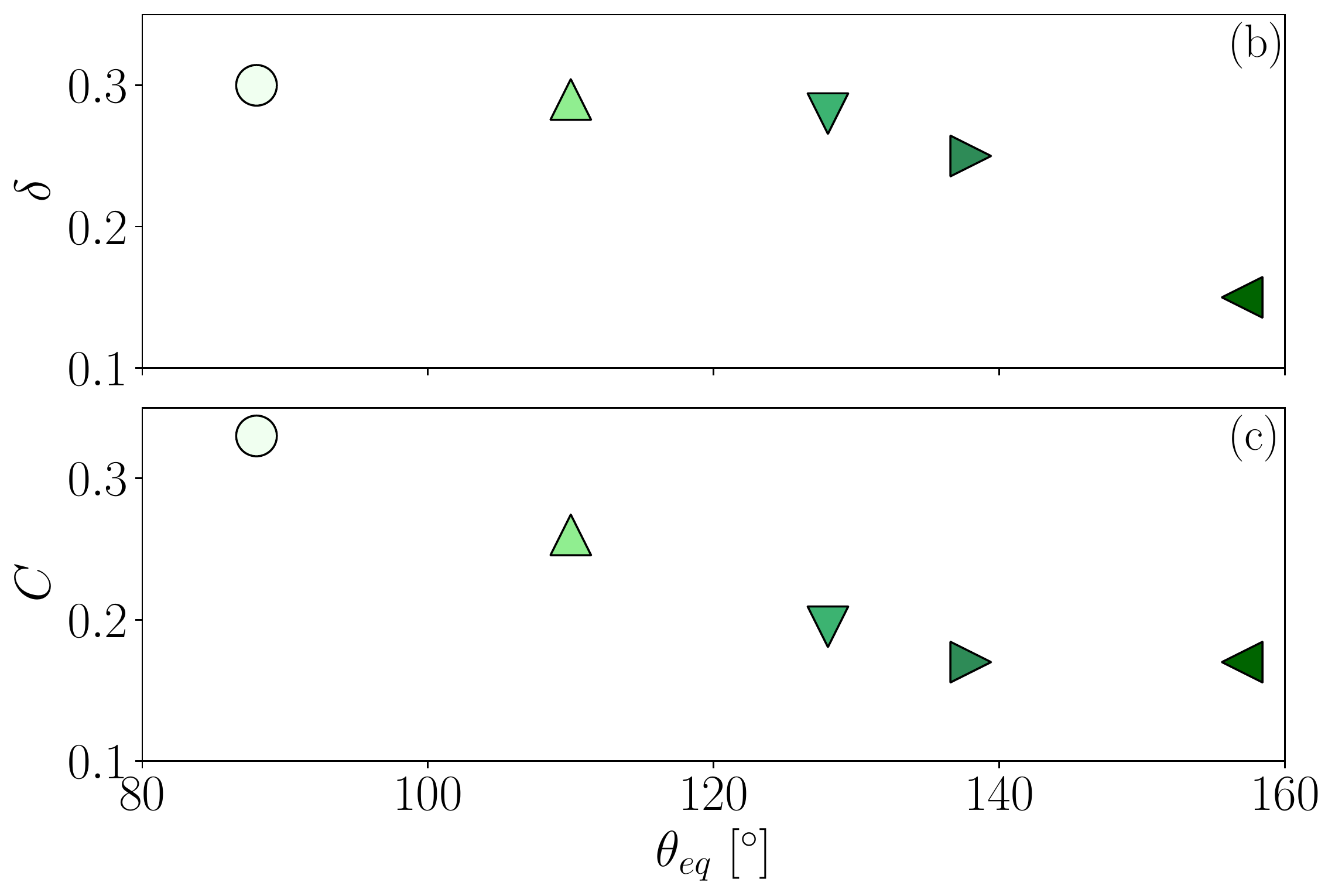}
    \end{tabular}
    \caption{Spreading experiment for a pure liquid droplet, $\alpha = (\alpha_1,0,0)$, on a flat surface. Panel (a): radius of the contact area $r$ as a function of time $t$, where $r$ and $t$ are reported normalised to the initial radius $R$ and the characteristic time $t^* = (\rho R^3 / \sigma)^{1/2}$, respectively. Different symbols/colours refer to different values of the equilibrium contact angle $\theta_{eq}$. In all cases, we observe a scaling law $r/R = C (t/t^*)^\delta$. The solid line indicates the scaling with $\delta = 3/10$, while the dotted line refers to scaling $\delta = 3/20$. Panels (b) and (c) show the value of the dimensionless exponent $\delta$ and the dimensionless prefactor $C$, respectively, as a function of $\theta_{eq}$.\label{fig:tanner's_law}
    }
\end{figure*}
To showcase the versatility of the interface model proposed in Sec.~\ref{sec:membranemodel}, we perform a double analysis by measuring the deformation of coated droplets and soft particles undergoing a shear flow. Indeed, on the one hand, we benchmark our model with what is known in the literature for the case of a pure droplet, while, on the other hand, we explore the different scenarios associated to each particle case listed in Table~\ref{table:cases}. In this setup, we run simulations for particles with an initial radius $R=19$ lbu, placed in a channel with a distance between the two walls H=128 lbu. The system has the same size along the other two directions, x and y. In order to vary the capillary number Ca, we systematically tune the values of $\bar{\alpha}$, keeping fixed the shear rate $\dot{\gamma}$ by the constraint of low Reynolds number (Re=$10^{-2}$). Without loss of generality, we set the fluid density $\rho=1$ lbu and the relaxation time $\tau=1$ lbu, resulting in a dynamic viscosity of the particle $\mu$=1/6 lbu. In Fig.~\ref{fig:deformation}(a) and (d), we report simulation data for the time-evolution of the deformation index defined as
\begin{equation}\label{eq:taylor_deform}
      D(t) = \dfrac{r_1(t)-r_3(t)}{r_1(t)+r_3(t)},
\end{equation}
where $r_1$ and $r_3$ are the main particle semi-axes in the shear plane (see the top of Fig.~\ref{fig:deformation} for a sketch). The simulation time is normalised with $\dot{\gamma}$. Results show a very good agreement between simulations and the time-evolution of the deformation $D$ defined in Eq.~\eqref{eq:deformation_J}, obtained from the analytical solutions of Eqs.~\eqref{eq:evJK} (dashed lines). In addition, the steady-state value of the deformation $D$ can be analytically estimated as a function of the triad of $\alpha$ as
\begin{equation}\label{eq:deformation}
D = \left[ \frac{5\alpha_1 \left(3 \alpha_{2} + 4 \alpha_{3}\right)}{4 \left(3 \alpha_{1} \alpha_{2} + 5 \alpha_{1} \alpha_{3} + 2 \alpha_{2} \alpha_{3} + 2 \alpha_{3}^{2}\right)}\right]\mbox{Ca},
\end{equation}
where Ca=$\mu R \dot{\gamma}/\alpha_1$. Since Eq.~\eqref{eq:deformation} has been computed with $\alpha_2, \ \alpha_3 \neq 0$, it does not hold for a pure droplet, for which $\alpha = \Bar{\alpha} (1,0,0)$. In the latter case we have~\cite{barthes1981time}
\begin{equation}\label{eq:deformation_droplet}
    D = \dfrac{19 \lambda + 16}{16 \lambda + 16}\mbox{Ca},
\end{equation}
with $\lambda =1$ in the present work. Fig.~\ref{fig:deformation}(b) and (e) confirm the agreement between simulation data and Eqs.~\eqref{eq:deformation} and ~\eqref{eq:deformation_droplet} in the limit of small deformations (i.e., small Ca), while it diverges for larger values of $D$. Note that in Fig.~\ref{fig:deformation}(b) the theoretical prediction for cases $\alpha = \Bar{\alpha} (1,0,1)$ and $\alpha = \Bar{\alpha} (1,1,1)$ are so close to not being distinguishable. 

In addition, to complete the picture of soft particle dynamics under shear flow, we report in Fig.~\ref{fig:deformation}(c) and (f) the inclination angle $\Theta$ (see top panel) as a function of the capillary number Ca. In the limit of small deformations and the case of a pure droplet with $\lambda =1$, these results are again in agreement with what is expected from simulations~\cite{Gounley16} (black crosses) and the theory of Chaffey and Brenner~\cite{CB_law_67} (dotted black line) which reads
\begin{equation}\label{eq:inclination_angle}
    \Theta = \dfrac{\pi}{4} - \dfrac{(19\lambda + 16)(2\lambda+3)}{80(\lambda+1)} \mbox{Ca}.
\end{equation}
For non-pre-stressed particle models, we observe a stronger dependency on Ca, probably due to the higher rigidity.\\
To summarize, we find a good agreement for the time-evolution of the particle deformation $D(t)$ and its steady-state value $D$ between the analytical solution of the model equations~\eqref{eq:evJK} and our numerical model for both coated droplets and soft particles. This is true in the limit of small deformation, which is the basic assumption behind the theory~\cite{barthes1981time}. Furthermore, in the case of a pure droplet, both $D$ and the inclination angle $\Theta$ follow the analytical predictions. It is worth noting that this benchmark contribute also to the validation of our generalised interface model against the interface response to an external flow.

\section{Wetting dynamics}
\begin{figure*}[t!]
    \centering
    \begin{tabular}{c}
 \includegraphics[width=.95\linewidth]{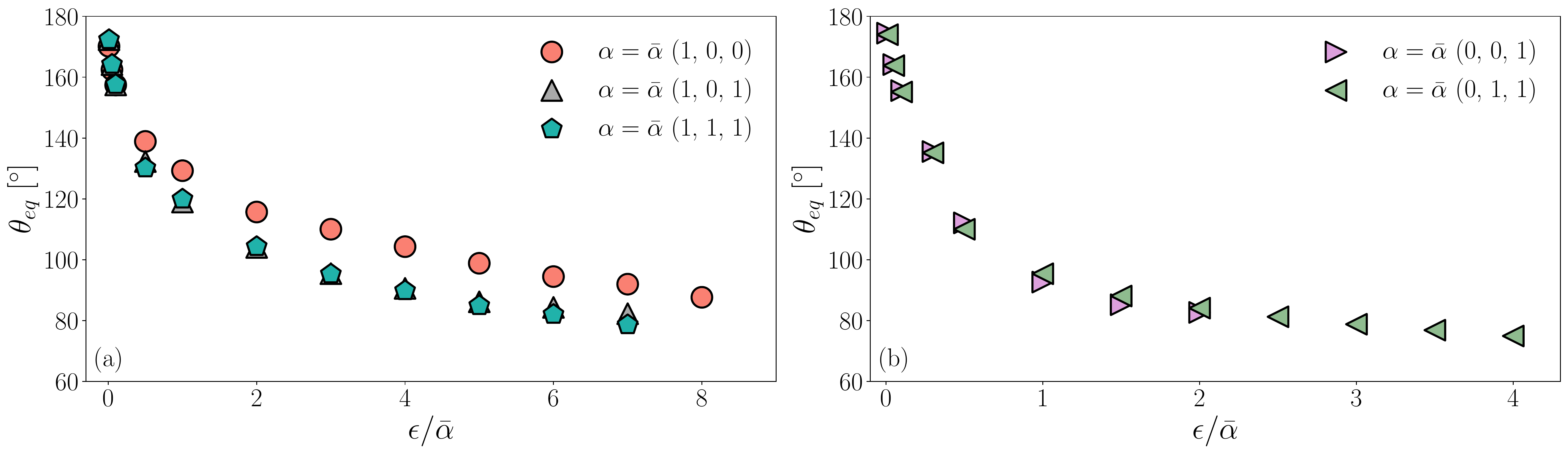}\\
 \includegraphics[width=.95\linewidth]{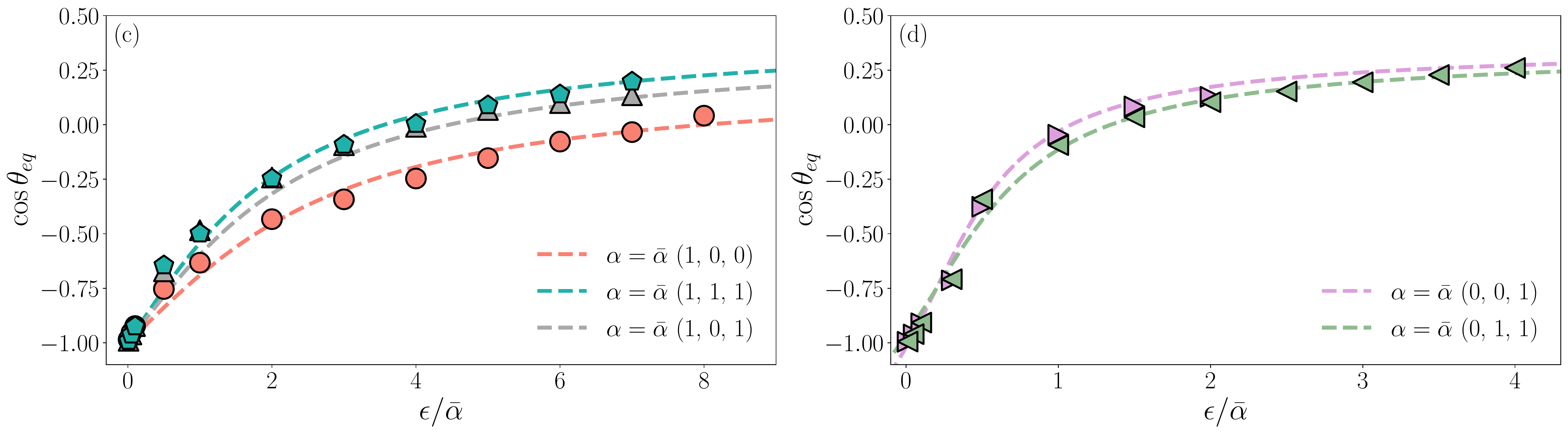}
 \end{tabular}
    \caption{Experiment of wetting dynamics. Panels (a) and (b): Equilibrium contact angle $\theta_{eq}$ as a function of the wall-particle interaction intensity $\epsilon$, normalised to $\bar{\alpha}$. Panels (c) and (d): Corresponding values of $\cos{\theta_{eq}}$. Left panels ((a) and (c)) refer to pre-stressed particle models, while right panels ((b) and (d)) refer to non-pre-stressed particle models. In all panels, different symbols/colours refer to different models, while dashed lines indicate fitting curves with Eq.~\eqref{eq:cos_theta_prediction} (values of fitting parameters are listed in Table~\ref{table:cases}). Data refer to simulations with a number of triangular faces equal to $N_f = 16820$.\label{fig:contact_angles}}
\end{figure*}
\subsection{Model validation}\label{sec:validation}
We now analyse the wetting dynamics of coated droplets and soft particles simulated using the interface model discussed in Sec.~\ref{sec:membranemodel}. In this kind of experiment, we consider a single particle with initial radius $R$ and placed close to a flat wall, i.e., its initial position is such that the z-coordinate of its centre-of-mass $\mbox{Z}_{\mathrm{\tiny CM}}$ is at a distance $R$ from the wall to let it feel the action of an attractive wall-interface interaction with intensity $\epsilon$ (see Eq.~\eqref{eq:LJ_interaction} and Fig.~\ref{fig:sketch} for a pictorial view). In our implementation, since we do not have direct control on $\sigma_{sl}$, $\sigma_{sg}$ appearing in Eq.~\eqref{eq:young_eq}, we consider $\epsilon$ playing the role of an effective solid surface tension as $\epsilon \propto \sigma_{sl}-\sigma_{sg}$.

Before entering into the details of the wetting dynamics of pre-stressed and non-pre-stressed particles, we validate our implementation by quantitatively investigating the spreading dynamics of a pure droplet,  $\alpha = \bar{\alpha}(1,0,0)$, by comparing the time evolution of the radius $r$ of the contact area with the literature. Indeed, it has been observed that this observable scales in time as
\begin{equation}\label{eq:tanner_law}
r = Ct^\delta,    
\end{equation}
where both the prefactor $C$ and the exponent $\delta$ can vary. When capillary forces drive the droplet spreading and inertial effects are negligible, Eq.~\eqref{eq:tanner_law} coincides with the Tanner's law~\cite{Tanner79}, predicting an exponent $\delta = 1/10$. Contrariwise, when capillary and inertial forces are balanced, it has been observed that the value of the exponent can vary with some factors, such as viscosity~\cite{LegendreMiglio13}, surface tension~\cite{LegendreMiglio13}, droplet initial shape~\cite{Courbin09} and wettability~\cite{Bird08,Carlson11,Winkels12,LegendreMiglio13}. In particular, a value of $\delta=1/2$ has been observed in the case of very small contact angles. In Fig.~\ref{fig:tanner's_law}(a) we report the time evolution of $r$, normalised to the initial radius $R$ at varying equilibrium contact angles $\theta_{eq}$. A scaling law following Eq.~\eqref{eq:tanner_law} is observed, with the exponent $\delta$ slightly decreases at increasing $\theta_{eq}$, in agreement with Ref.~\cite{Bird08} (see Fig.~\ref{fig:tanner's_law}(b)). This implies that in our simulations of wetting dynamics, the inertia is not negligible and it plays a role in resisting the deformation. Note that, as later highlighted in Fig.~\ref{fig:contact_angles}(a), our model can capture only cases of large contact angles ($88^{\circ} \leq \theta_{eq} \leq 180^{\circ}$), indeed $\delta$ approaches but does not reach a value of close to 1/2 which is characteristic for small contact angles~\cite{Bird08,Carlson11,Winkels12,LegendreMiglio13}. This is comparable, for example, to the case of a liquid metal droplet, which has been observed to never assume values of $\theta_{eq} < 100^{\circ}$ also in the case of oxidization, suggesting that our model can be used to study such kind of system. Concerning the prefactor $C$ (Fig.~\ref{fig:tanner's_law}(c)), it decreases as $\theta_{eq}$ increases, once again in agreement with Ref.~\cite{Bird08}. Note that the ``jumps'' in $r$ that are visible in Fig.~\ref{fig:tanner's_law}(a) for $\theta_{eq} = 157^{\circ}$ originate from the numerical error in measuring very small variations of the contact area.

\subsection{Results}\label{sec:results}
\begin{figure*}[t!]
    \centering
    \includegraphics[width=.95\linewidth]{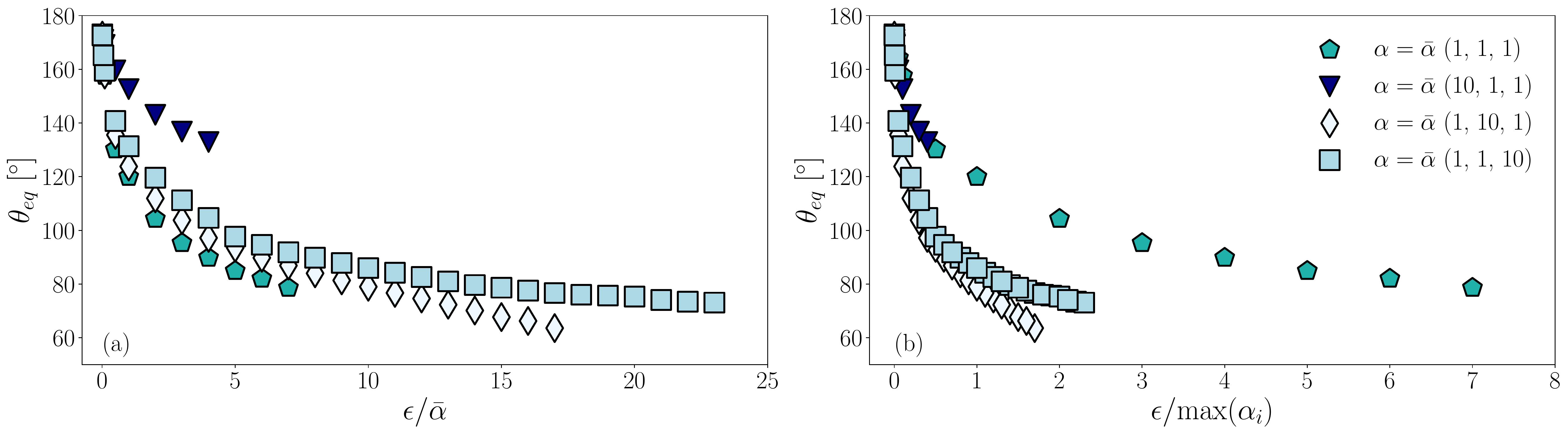}
    \caption{Equilibrium contact angle $\theta_{eq}$ as a function of the wall-particle interaction intensity $\epsilon$, normalised to $\bar{\alpha}$ (panel (a)) and max$( \alpha_i)$ with $i=1,2,3$ (panel (b)). All data refer to the case with all components of $\alpha = (\alpha_1,\alpha_2,\alpha_3)$ turned on, but different symbols/colours refer to a different ``extreme'' cases, where one of the three parameters is increased by an order of magnitude. The number of triangular faces $N_f$ is the same as for the data in Fig.~\ref{fig:contact_angles}.\label{fig:estreme_cases}}
\end{figure*}
With the aim of simulating the wetting dynamics of droplets with complex interface properties, we explore both pre-stressed ($\alpha_1 > 0$) and non-pre-stressed ($\alpha_1 = 0$) particles. For each system, characterized by $\alpha = (\alpha_1,\alpha_2,\alpha_3) = \bar{\alpha} (0/1,0/1,0/1)$, we apply the same strategy used for the benchmark, which we summarize again for the sake of clarity. First, we fix the value of $\bar{\alpha}$ to $10^{-4}$ lbu. This choice fixes both the surface tension for pre-stressed particles and the strain modulus for non-pre-stressed particles. Then we measure $\theta_{eq}$ as a function of the wall-particle interaction energy $\epsilon$. In  Appendix~\ref{app:resolution} we discuss also a resolution test for the wetting dynamics.\\
In Fig.~\ref{fig:contact_angles}(a) and (b) we report  the measured $\theta_{eq}$ as a function of the ratio 
$\epsilon/\bar{\alpha}$ for pre-stressed and non-pre-stressed particle models, respectively. In Fig.~\ref{fig:contact_angles}(c) and (d) we report for convenience the corresponding values of $\cos{\theta_{eq}}$. After the largest value of $\epsilon/\bar{\alpha}$ reported in each plot, numerical instabilities appear in the contact area region; these set the limit of applicability of our approach in terms of contact angles that can be modeled.
Concerning pre-stressed particles, pure droplet (circles) appear to be marginally more stable with respect to the choice of $\epsilon$ than the other particles (softly coated particles, $\alpha = \bar{\alpha} (1,0,1)$, upward triangles; rigidly coated particles, $\alpha = \bar{\alpha} (1,1,1)$, pentagons), but can reach only a slightly higher contact angle  ($\theta_{eq}=88$ vs. $\theta_{eq} = 79^{\circ}$). Notice that the cases $\alpha = \bar{\alpha} (1,0,1)$ and $\alpha = \bar{\alpha} (1,1,1)$ are very similar, meaning that when the system is very rigid, the dilatational contribution given by $\alpha_2$ is not relevant for the equilibrium contact angle $\theta_{eq}$. This result is in contrast with what we observed in the shear flow. Non-pre-stressed particle models (Fig.~\ref{fig:contact_angles}(b) and (d)) are stable for a more limited range of values of $\epsilon/\bar{\alpha}$. After around the value of $\epsilon/\bar{\alpha} = 1.5$ the contact angle does not drop significantly anymore. Similarly to the pre-stressed case, $\alpha_2$ does not seem to have any influence on $\theta_{eq}$. The behaviour of $\cos{\theta_{eq}}$ as a function of $\epsilon/\bar{\alpha}$ follows very well the empirical behavior
\begin{equation}\label{eq:cos_theta_prediction}
    \cos{\theta_{eq}} \simeq a \tan^{-1}{\left[b (\epsilon/\bar{\alpha}-c)\right]} - d
\end{equation}
where $a$, $b$, $c$ and $d$ are fitting parameters depending on the type of particle (see Table~\ref{table:cases} and dashed lines in Fig.~\ref{fig:contact_angles}(c) and (d)). Eq.~\eqref{eq:cos_theta_prediction} differs from Eq.~\eqref{eq:young_eq} because, as mentioned above, the model we present in this work misses the direct control of the wall surface tensions but rather drives the mechanical interaction between the particle and the solid surface. Obviously, the fitting constants represent (unknown) functions of the parameters $\alpha_1$, $\alpha_2$, $\alpha_3$ and $\epsilon$. By increasing separately by a factor of ten each of the components of $\alpha$, as reported in Fig.~\ref{fig:estreme_cases}(a), we can understand that the leading order behavior is dictated by $\alpha_1$, while $\alpha_2$ and $\alpha_3$ provide relatively minor changes in $\theta_{eq}$. In addition, from Fig.~\ref{fig:estreme_cases}(b) one can see that $\alpha_1$ must enter in Eq.~\eqref{eq:cos_theta_prediction} in the ratio with $\epsilon$, because data for the increased $\alpha_1$ (downward triangles) collapses onto the original case (pentagons) once plotted as a function $\epsilon/\alpha_1$. The remaining parameters $\alpha_2$ and $\alpha_3$, instead, do not appear to provide a similar scaling, thus meaning that these two parameters may functionally enter in the other fitting parameters of Eq.~\eqref{eq:cos_theta_prediction}.
\begin{figure*}[t!]
    \centering
    \begin{tabular}{c}
\includegraphics[width=.8\linewidth]{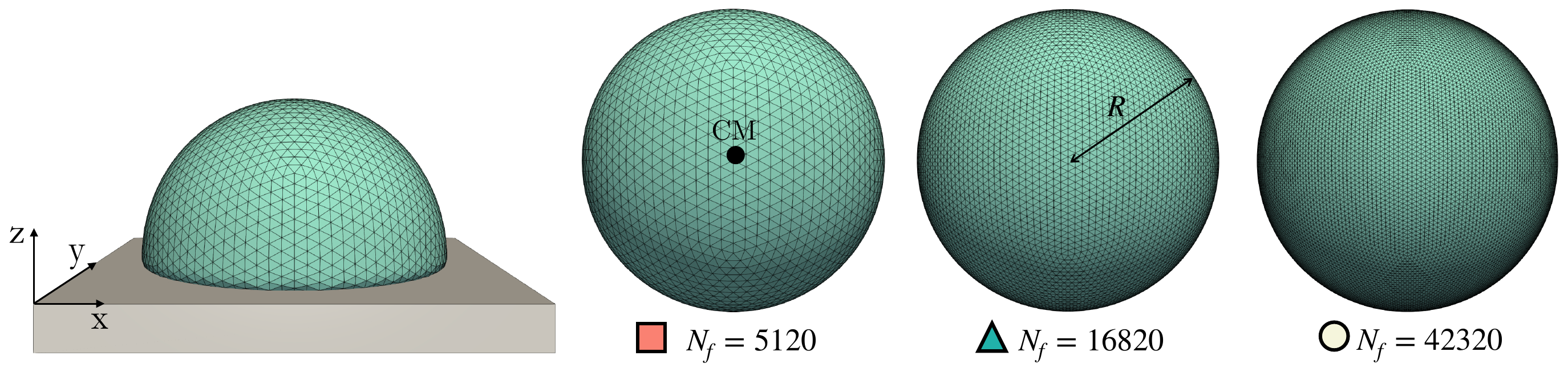} 
\end{tabular}
\begin{tabular}{c c}
 \includegraphics[width=.5\linewidth]{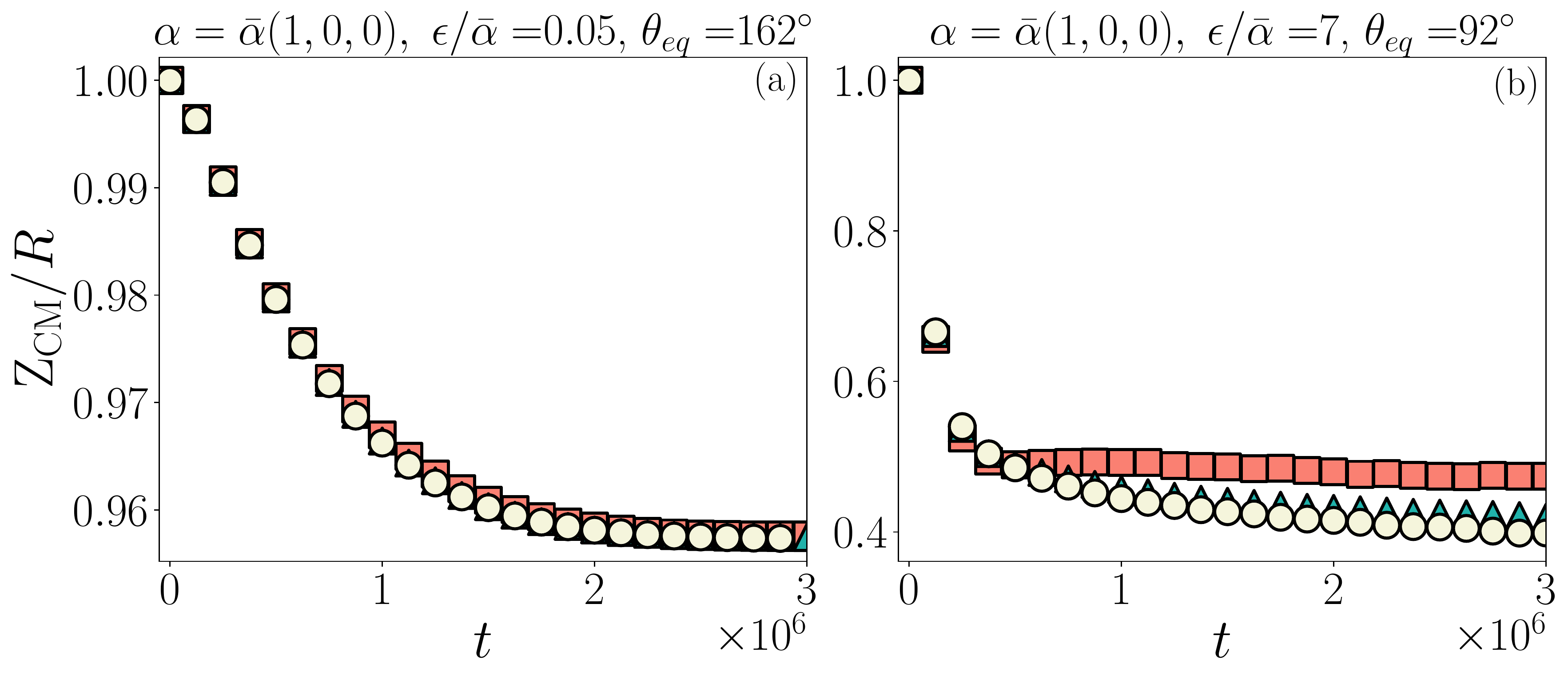} &
  \includegraphics[width=.5\linewidth]{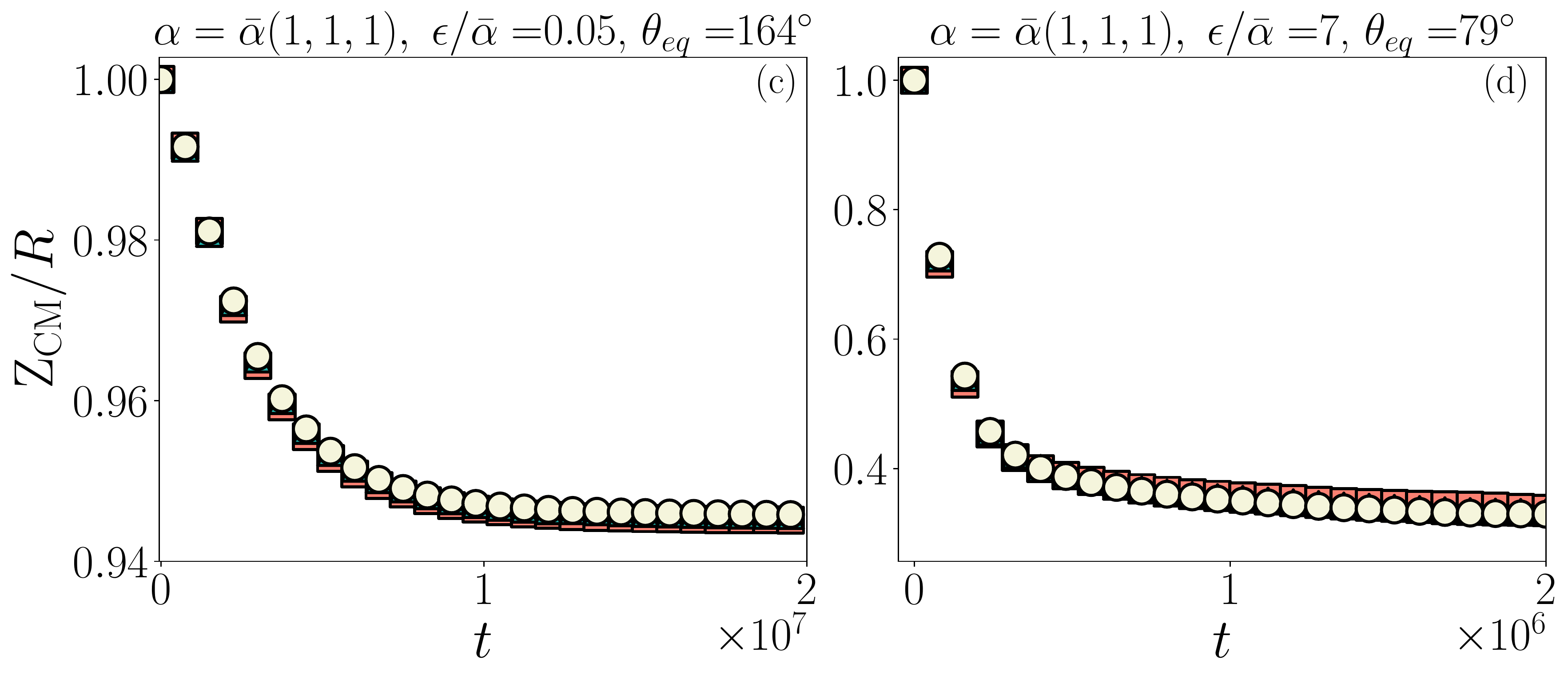}
 \end{tabular}
    \caption{Resolution test for the wetting dynamics experiment a pure droplet with $\alpha = \bar{\alpha} (1,0,0)$ (panels (a) and (b)) and for a mixed system with $\alpha = \bar{\alpha} (1,1,1)$ (panels (c) and (d)). We compare the time evolution of the z-coordinate of the centre-of-mass $\mbox{Z}_{\mathrm{\tiny CM}}$ for different resolutions, given in terms of the number of mesh triangular faces $N_f$. Time $t$ is shown in simulation units. \label{fig:resolution_test}}
\end{figure*}

\section{Conclusions}\label{sec:conclusions}
In this work we introduced a novel numerical framework to accurately characterize coated droplets and soft particles. This approach is based on the thory of Barth\`es-Biesel and Rallison~\cite{barthes1981time} and enables us to capture the unique behavior that is intermediate between that of a pure droplet and a capsule. With this generalised constitutive law we are able to capture the special properties of a wide spectrum of coated droplets, for example, liquid metal droplets surrounded by an oxide layer.
In the present approach, the interface strain energy is written in terms of three parameters that play the role of material properties, i.e., the pre-stress ($\alpha_1$), the resistance against area dilatation ($\alpha_2$), and the resistance against shear deformation ($\alpha_3$). With the choice of these three parameters, we explore different types of coated droplets, from pure liquid droplets to soft particles. We validate our methodology with the theoretical predictions and recent experiments in both shear flow and wetting experiments, and we explore the limits of the model in terms of $\alpha_{1,2,3}$. We plan to enrich this description by including new contributions to the presented model, for example, to mimic the thickness of the oxide layer in the case of liquid metal droplets. The latter could be useful to mimic the dynamics of other complex droplets, such as liquid marbles.

\section*{Author declarations}
The authors have no conflicts to disclose.

\begin{acknowledgments}
This work has received financial support from the Deutsche Forschungsgemeinschaft (DFG, German Research Foundation) – Project-ID 431791331 – SFB 1452 ``Catalysis at liquid interfaces'' and research unit FOR2688 ``Instabilities, Bifurcations and Migration in Pulsatile Flows'' (Project-ID 417989464).  
This work was supported by the Italian Ministry of University and Research (MUR) under the FARE programme, project ``Smart-HEART''.
The authors gratefully acknowledge the Gauss Centre for Supercomputing e.V. (\url{www.gauss-centre.eu}) for funding this project by providing computing time through the John von Neumann Institute for Computing (NIC) on the GCS Supercomputer JUWELS (\cite{JUWELS}) at Jülich Supercomputing Centre (JSC).
\end{acknowledgments}

\section*{Data Availability}

The data that support the findings of this study are available from the corresponding author upon reasonable request.

\appendix

\section{Resolution test for wetting dynamics}\label{app:resolution}
Results shown in Figs.~\ref{fig:contact_angles} and~\ref{fig:estreme_cases} required a test to choose the best resolution in terms of accuracy and computational effort. In Fig.~\ref{fig:resolution_test} we show the time evolution of the z-coordinate of its centre-of-mass $\mbox{Z}_{\mathrm{\tiny CM}}$, normalised by the initial radius $R$, for a pure droplet case, $\alpha = \bar{\alpha} (1,0,0)$, shown in Fig.~\ref{fig:resolution_test}(a) and (b), and a rigidly coated droplet,  $\alpha = \bar{\alpha} (1,1,1)$, shown in  Fig.~\ref{fig:resolution_test}(c) and (d). We report two values of $\epsilon/\bar{\alpha}$, i.e., 0.05, (Fig.~\ref{fig:resolution_test}(a) and (c)), and 7 (Fig.~\ref{fig:resolution_test}(b) and (d)), resulting in a large and a small equilibrium contact angle for both systems. As long as the contact angle is very large then all resolutions are equivalent. However, moving towards $\theta_{eq} \sim 80^{\circ}$, a large number of $N_f$ is required for more precise contact angle measurements in the case of a pure droplet. The latter statement follows from the way we compute $\theta_{eq}$, i.e., by fitting the droplet shape with a circumference cut by a chord (i.e., the wall). To perform the fitting procedure, we take a slice of the particle mesh involving a number of nodes which is roughly $\frac{4}{\sqrt[4]{3}}\sqrt{\pi N_f}$. Furthermore, simulations with $N_f = 42320$ show the same dynamics as $N_f = 16820$ but they require a higher computational cost, leading to the choice made to produce the data reported in Figs.~\ref{fig:contact_angles} and~\ref{fig:estreme_cases}, to run simulations with $N_f = 16820$.


\bibliographystyle{ieeetr}
\bibliography{biblio}

\begin{thebibliography}{10}

\bibitem{DeGennes85}
P.-G. De~Gennes, ``Wetting: statics and dynamics,'' {\em Reviews of modern
  physics}, vol.~57, p.~827, 1985.

\bibitem{Brochard92}
F.~Brochard-Wyart and P.~De~Gennes, ``Dynamics of partial wetting,'' {\em
  Advances in colloid and interface science}, vol.~39, pp.~1--11, 1992.

\bibitem{Bonn01}
D.~Bonn, ``Wetting transitions,'' {\em Current opinion in colloid \& interface
  science}, vol.~6, pp.~22--27, 2001.

\bibitem{DeConinck01}
J.~De~Coninck, M.~J. de~Ruijter, and M.~Vou{\'e}, ``Dynamics of wetting,'' {\em
  Current opinion in colloid \& interface science}, vol.~6, pp.~49--53, 2001.

\bibitem{Young1805}
T.~Young, ``{III.} an essay on the cohesion of fluids,'' {\em Philosophical
  transactions of the {R}oyal {S}ociety of {L}ondon}, vol.~95, pp.~65--87,
  1805.

\bibitem{andreotti2020statics}
B.~Andreotti and J.~H. Snoeijer, ``Statics and dynamics of soft wetting,'' {\em
  Annual review of fluid mechanics}, vol.~52, pp.~285--308, 2020.

\bibitem{de2008wetting}
J.~De~Coninck and T.~D. Blake, ``Wetting and molecular dynamics simulations of
  simple liquids,'' {\em Annu. Rev. Mater. Res.}, vol.~38, pp.~1--22, 2008.

\bibitem{bonn2009wetting}
D.~Bonn, J.~Eggers, J.~Indekeu, J.~Meunier, and E.~Rolley, ``Wetting and
  spreading,'' {\em Reviews of modern physics}, vol.~81, no.~2, p.~739, 2009.

\bibitem{snoeijer2013moving}
J.~H. Snoeijer and B.~Andreotti, ``Moving contact lines: scales, regimes, and
  dynamical transitions,'' {\em Annual review of fluid mechanics}, vol.~45,
  pp.~269--292, 2013.

\bibitem{Taccardi17}
N.~Taccardi, M.~Grabau, J.~Debuschewitz, M.~Distaso, M.~Brandl, R.~Hock,
  F.~Maier, C.~Papp, J.~Erhard, C.~Neiss, {\em et~al.}, ``Gallium-rich {Pd--Ga}
  phases as supported liquid metal catalysts,'' {\em Nature chemistry}, vol.~9,
  pp.~862--867, 2017.

\bibitem{Hofer23}
A.~Hofer, N.~Taccardi, M.~Moritz, C.~Wichmann, S.~H{\"u}bner, D.~Drobek,
  M.~Engelhardt, G.~Papastavrou, E.~Spiecker, C.~Papp, {\em et~al.},
  ``Preparation of geometrically highly controlled ga particle arrays on
  quasi-planar nanostructured surfaces as a scalms model system,'' {\em RSC
  advances}, vol.~13, pp.~4011--4018, 2023.

\bibitem{Daeneke18}
T.~Daeneke, K.~Khoshmanesh, N.~Mahmood, I.~A. De~Castro, D.~Esrafilzadeh,
  S.~Barrow, M.~Dickey, and K.~Kalantar-Zadeh, ``Liquid metals: fundamentals
  and applications in chemistry,'' {\em Chemical Society Reviews}, vol.~47,
  pp.~4073--4111, 2018.

\bibitem{Allioux22}
F.-M. Allioux, M.~B. Ghasemian, W.~Xie, A.~P. O'Mullane, T.~Daeneke, M.~D.
  Dickey, and K.~Kalantar-Zadeh, ``Applications of liquid metals in
  nanotechnology,'' {\em Nanoscale Horizons}, vol.~7, pp.~141--167, 2022.

\bibitem{Jeyakumar11}
M.~Jeyakumar, M.~Hamed, and S.~Shankar, ``Rheology of liquid metals and
  alloys,'' {\em Journal of non-Newtonian fluid mechanics}, vol.~166,
  pp.~831--838, 2011.

\bibitem{Doudrick14}
K.~Doudrick, S.~Liu, E.~M. Mutunga, K.~L. Klein, V.~Damle, K.~K. Varanasi, and
  K.~Rykaczewski, ``Different shades of oxide: From nanoscale wetting
  mechanisms to contact printing of gallium-based liquid metals,'' {\em
  Langmuir}, vol.~30, pp.~6867--6877, 2014.

\bibitem{Joshipura21}
I.~D. Joshipura, K.~A. Persson, V.~K. Truong, J.-H. Oh, M.~Kong, M.~H. Vong,
  C.~Ni, M.~Alsafatwi, D.~P. Parekh, H.~Zhao, {\em et~al.}, ``Are contact angle
  measurements useful for oxide-coated liquid metals?,'' {\em Langmuir},
  vol.~37, pp.~10914--10923, 2021.

\bibitem{zheng2022liquid}
L.~Zheng, S.~Handschuh-Wang, Z.~Ye, and B.~Wang, ``Liquid metal droplets
  enabled soft robots,'' {\em Applied Materials Today}, vol.~27, p.~101423,
  2022.

\bibitem{Bormashenko11}
E.~Bormashenko, ``Liquid marbles: properties and applications,'' {\em Current
  Opinion in Colloid \& Interface Science}, vol.~16, pp.~266--271, 2011.

\bibitem{Ni2021}
E.~Ni, T.~Li, Y.~Ruan, Y.~Ma, Y.~Wang, Y.~Jiang, and H.~Li, ``Modeling of
  wetting transition of liquid metals on organic liquid surfaces,'' {\em
  Langmuir}, vol.~37, pp.~9429--9438, 2021.

\bibitem{Zhao10}
Y.~Zhao, J.~Fang, H.~Wang, X.~Wang, and T.~Lin, ``Magnetic liquid marbles:
  manipulation of liquid droplets using highly hydrophobic fe3o4
  nanoparticles,'' {\em Advanced materials}, vol.~22, pp.~707--710, 2010.

\bibitem{Avruamescu18}
R.-E. Avr{\u{a}}mescu, M.-V. Ghica, C.~Dinu-P{\^\i}rvu, D.~I. Udeanu, and
  L.~Popa, ``Liquid marbles: From industrial to medical applications,'' {\em
  Molecules}, vol.~23, p.~1120, 2018.

\bibitem{Biferale07}
L.~Biferale, R.~Benzi, M.~Sbragaglia, S.~Succi, and F.~Toschi,
  ``Wetting/dewetting transition of two-phase flows in nano-corrugated
  channels,'' {\em Journal of computer-aided materials design}, vol.~14,
  pp.~447--456, 2007.

\bibitem{Yan07}
Y.~Yan and Y.~Zu, ``A lattice {B}oltzmann method for incompressible two-phase
  flows on partial wetting surface with large density ratio,'' {\em Journal of
  Computational Physics}, vol.~227, pp.~763--775, 2007.

\bibitem{HuangSukop07}
H.~Huang, D.~T. Thorne~Jr, M.~G. Schaap, and M.~C. Sukop, ``Proposed
  approximation for contact angles in {S}han-and-{C}hen-type multicomponent
  multiphase lattice {B}oltzmann models,'' {\em Physical Review E}, vol.~76,
  p.~066701, 2007.

\bibitem{Attar09}
E.~Attar and C.~K{\"o}rner, ``Lattice {B}oltzmann method for dynamic wetting
  problems,'' {\em Journal of colloid and interface science}, vol.~335,
  pp.~84--93, 2009.

\bibitem{Tanaka11}
Y.~Tanaka, Y.~Washio, M.~Yoshino, and T.~Hirata, ``Numerical simulation of
  dynamic behavior of droplet on solid surface by the two-phase lattice
  {B}oltzmann method,'' {\em Computers \& fluids}, vol.~40, pp.~68--78, 2011.

\bibitem{HKH11}
J.~Hyv\"aluoma, C.~Kunert, and J.~Harting, ``Simulations of slip flow on
  nanobubble-laden surfaces,'' {\em Journal of Physics: Condensed Matter},
  vol.~23, p.~184106, 2011.

\bibitem{Jansen13}
H.~P. Jansen, K.~Sotthewes, J.~van Swigchem, H.~J. Zandvliet, and E.~S. Kooij,
  ``Lattice {B}oltzmann modeling of directional wetting: Comparing simulations
  to experiments,'' {\em Physical Review E}, vol.~88, p.~013008, 2013.

\bibitem{Wang17}
L.~Wang and J.~Sun, ``The application of axisymmetric lattice {B}oltzmann
  two-phase model on simulations of liquid film dewetting,'' {\em Journal of
  Applied Physics}, vol.~122, p.~085305, 2017.

\bibitem{akai2018wetting}
T.~Akai, B.~Bijeljic, and M.~J. Blunt, ``Wetting boundary condition for the
  color-gradient lattice boltzmann method: Validation with analytical and
  experimental data,'' {\em Advances in Water Resources}, vol.~116, pp.~56--66,
  2018.

\bibitem{Zitz19}
S.~Zitz, A.~Scagliarini, S.~Maddu, A.~A. Darhuber, and J.~Harting, ``Lattice
  {B}oltzmann method for thin-liquid-film hydrodynamics,'' {\em Phys. Rev. E},
  vol.~100, p.~033313, Sep 2019.

\bibitem{Wang20}
X.~Wang, B.~Xu, Y.~Wang, and Z.~Chen, ``Directional migration of single droplet
  on multi-wetting gradient surface by 3d lattice {B}oltzmann method,'' {\em
  Computers \& Fluids}, vol.~198, p.~104392, 2020.

\bibitem{Pelusi22}
F.~Pelusi, M.~Sega, and J.~Harting, ``Liquid film rupture beyond the thin-film
  equation: A multi-component lattice {B}oltzmann study,'' {\em Physics of
  Fluids}, vol.~34, p.~062109, 2022.

\bibitem{ShanChen93}
X.~Shan and H.~Chen, ``Lattice {B}oltzmann model for simulating flows with
  multiple phases and components,'' {\em Physical review E}, vol.~47, p.~1815,
  1993.

\bibitem{Swift96}
M.~R. Swift, E.~Orlandini, W.~Osborn, and J.~Yeomans, ``Lattice {B}oltzmann
  simulations of liquid-gas and binary fluid systems,'' {\em Physical Review
  E}, vol.~54, no.~5, p.~5041, 1996.

\bibitem{barthes1981time}
D.~Barthès-Biesel and J.~Rallison, ``The time-dependent deformation of a
  capsule freely suspended in a linear shear flow,'' {\em Journal of Fluid
  Mechanics}, vol.~113, pp.~251--267, 1981.

\bibitem{skalakStrainEnergyFunction1973}
R.~Skalak, A.~Tozeren, R.~Zarda, and S.~Chien, ``Strain {{Energy Function}} of
  {{Red Blood Cell Membranes}},'' {\em Biophysical Journal}, vol.~13,
  pp.~245--264, Mar. 1973.

\bibitem{matsunaga2016rheology}
D.~Matsunaga, Y.~Imai, T.~Yamaguchi, and T.~Ishikawa, ``Rheology of a dense
  suspension of spherical capsules under simple shear flow,'' {\em Journal of
  Fluid Mechanics}, vol.~786, pp.~110--127, 2016.

\bibitem{BarthesSgaier85}
D.~Barthes-Biesel and H.~Sgaier, ``Role of membrane viscosity in the
  orientation and deformation of a spherical capsule suspended in shear flow,''
  {\em Journal of Fluid Mechanics}, vol.~160, pp.~119--135, 1985.

\bibitem{benzi1992lattice}
R.~Benzi, S.~Succi, and M.~Vergassola, ``The lattice {B}oltzmann equation:
  theory and applications,'' {\em Physics Reports}, vol.~222, pp.~145--197,
  1992.

\bibitem{Kruger16}
T.~Krüger, H.~Kusumaatmaja, A.~Kuzmin, O.~Shardt, G.~Silva, and E.~M. Viggen,
  {\em The Lattice {{Boltzmann}} Method -- Principles and Practice}.
\newblock {Springer}, 2016.

\bibitem{Guo02}
Z.~Guo, C.~Zheng, and B.~Shi, ``Discrete lattice effects on the forcing term in
  the lattice {B}oltzmann method,'' {\em Physical Review E}, vol.~65,
  p.~046308, 2002.

\bibitem{Peskin02}
C.~S. Peskin, ``The immersed boundary method,'' {\em Acta numerica}, vol.~11,
  pp.~479--517, 2002.

\bibitem{Kaoui11}
B.~Kaoui, J.~Harting, and C.~Misbah, ``Two-dimensional vesicle dynamics under
  shear flow: effect of confinement,'' {\em Physical Review E}, vol.~83,
  p.~066319, 2011.

\bibitem{kruger2011efficient}
T.~Kr{\"u}ger, F.~Varnik, and D.~Raabe, ``Efficient and accurate simulations of
  deformable particles immersed in a fluid using a combined immersed boundary
  lattice {B}oltzmann finite element method,'' {\em Computers \& Mathematics
  with Applications}, vol.~61, pp.~3485--3505, 2011.

\bibitem{aouaneStructureRheologySuspensions2021}
O.~Aouane, A.~Scagliarini, and J.~Harting, ``Structure and rheology of
  suspensions of spherical strain-hardening capsules,'' {\em Journal of Fluid
  Mechanics}, vol.~911, p.~A11, Mar. 2021.

\bibitem{BAHK21}
C.~Bielinski, O.~Aouane, J.~Harting, and B.~Kaoui, ``Squeezing multiple soft
  particles into a constriction: transition to clogging,'' {\em Physical Review
  E}, vol.~104, p.~065101, 2021.

\bibitem{KKH14}
T.~Krüger, B.~Kaoui, and J.~Harting, ``Interplay of inertia and deformability
  on rheological properties of a suspension of capsules,'' {\em The Journal of
  Fluid Mechanics}, vol.~751, pp.~725--745, 2014.

\bibitem{guglietta2023suspensions}
F.~Guglietta, F.~Pelusi, M.~Sega, O.~Aouane, and J.~Harting, ``Suspensions of
  viscoelastic capsules: effect of membrane viscosity on transient dynamics,''
  {\em arXiv preprint arXiv:2302.03546}, 2023.

\bibitem{krugerDeformabilitybasedRedBlood2014}
T.~Kr{\"u}ger, D.~Holmes, and P.~V. Coveney, ``Deformability-based red blood
  cell separation in deterministic lateral displacement devices\textemdash{{A}}
  simulation study,'' {\em Biomicrofluidics}, vol.~8, p.~054114, Oct. 2014.

\bibitem{kaouiHowDoesConfinement2012}
B.~Kaoui, T.~Kr{\"u}ger, and J.~Harting, ``How does confinement affect the
  dynamics of viscous vesicles and red blood cells?,'' {\em Soft Matter},
  vol.~8, p.~9246, 2012.

\bibitem{liSimilarDistinctRoles2021}
P.~Li and J.~Zhang, ``Similar but {{Distinct Roles}} of {{Membrane}} and
  {{Interior Fluid Viscosities}} in {{Capsule Dynamics}} in {{Shear Flows}},''
  {\em Cardiovascular Engineering and Technology}, vol.~12, pp.~232--249, Apr.
  2021.

\bibitem{guglietta2020lattice}
F.~Guglietta, M.~Behr, L.~Biferale, G.~Falcucci, and M.~Sbragaglia, ``Lattice
  {B}oltzmann simulations on the tumbling to tank-treading transition: effects
  of membrane viscosity,'' {\em Philosophical Transactions of the Royal Society
  A: Mathematical, Physical and Engineering Sciences}, vol.~379, p.~20200395,
  2021.

\bibitem{guglietta2021loading}
F.~Guglietta, M.~Behr, G.~Falcucci, and M.~Sbragaglia, ``Loading and relaxation
  dynamics of a red blood cell,'' {\em Soft Matter}, vol.~17, pp.~5978--5990,
  2021.

\bibitem{liFiniteDifferenceMethod2019}
P.~Li and J.~Zhang, ``A finite difference method with subsampling for immersed
  boundary simulations of the capsule dynamics with viscoelastic membranes,''
  {\em International Journal for Numerical Methods in Biomedical Engineering},
  p.~e3200, Apr. 2019.

\bibitem{Guglietta2020}
F.~Guglietta, M.~Behr, L.~Biferale, G.~Falcucci, and M.~Sbragaglia, ``On the
  effects of membrane viscosity on transient red blood cell dynamics,'' {\em
  Soft Matter}, vol.~16, pp.~6191--6205, 2020.

\bibitem{taglientiReducedModelDroplet2023}
D.~Taglienti, F.~Guglietta, and M.~Sbragaglia, ``Reduced model for droplet
  dynamics in shear flows at finite capillary numbers,'' {\em Physical Review
  Fluids}, vol.~8, p.~013603, Jan. 2023.

\bibitem{KrugerPhDthesis12}
T.~Krüger, {\em Computer {{Simulation Study}} of {{Collective Phenomena}} in
  {{Dense Suspensions}} of {{Red Blood Cells}} under {{Shear}}}.
\newblock {Vieweg+Teubner Verlag}, 2012.

\bibitem{martys1996}
N.~S. Martys and H.~Chen, ``Simulation of multicomponent fluids in complex
  three-dimensional geometries by the lattice {B}oltzmann method,'' {\em
  Physical review E}, vol.~53, no.~1, p.~743, 1996.

\bibitem{HKH06}
J.~Harting, C.~Kunert, and H.~J. Herrmann, ``Lattice {B}oltzmann simulations of
  apparent slip in hydrophobic microchannels,'' {\em Europhysics Letters},
  vol.~75, p.~328, 2006.

\bibitem{li2014contact}
Q.~Li, K.~Luo, Q.~Kang, and Q.~Chen, ``Contact angles in the pseudopotential
  lattice {B}oltzmann modeling of wetting,'' {\em Physical Review E}, vol.~90,
  no.~5, p.~053301, 2014.

\bibitem{koplik1995}
J.~Koplik and J.~Banavar, ``Continuum deductions from molecular
  hydrodynamics,'' {\em Annual Reviews of Fluid Dynamics}, vol.~27, p.~257,
  1995.

\bibitem{koplik2000}
J.~Koplik and J.~Banavar, ``Molecular simulation of dewetting,'' {\em Physical
  Review Letters}, vol.~84, p.~4401, 2000.

\bibitem{koplik2006}
J.~Koplik, T.~Lo, M.~Rauscher, and S.~Dietrich, ``Pearling instability of
  nanoscale fluid flow confined to a chemical channel,'' {\em Physics of
  Fluids}, vol.~18, p.~0321045, 2006.

\bibitem{Semiromi11}
D.~T. Semiromi and A.~Azimian, ``Molecular dynamics simulation of nonodroplets
  with the modified {L}ennard-{J}ones potential function,'' {\em Heat and mass
  transfer}, vol.~47, pp.~579--588, 2011.

\bibitem{DRKHD12}
F.~D\"orfler, M.~Rauscher, J.~Koplik, J.~Harting, and S.~Dietrich, ``Micro- and
  nanoscale fluid flow on chemical channels,'' {\em Soft Matter}, vol.~8,
  p.~9221, 2012.

\bibitem{Gounley16}
J.~Gounley, G.~Boedec, M.~Jaeger, and M.~Leonetti, ``Influence of surface
  viscosity on droplets in shear flow,'' {\em Journal of Fluid Mechanics},
  vol.~791, p.~464–494, 2016.

\bibitem{CB_law_67}
C.~E. Chaffey and H.~Brenner, ``A second-order theory for shear deformation of
  drops,'' {\em Journal of Colloid and Interface Science}, vol.~24,
  pp.~258--269, 1967.

\bibitem{Tanner79}
L.~Tanner, ``The spreading of silicone oil drops on horizontal surfaces,'' {\em
  Journal of Physics D: Applied Physics}, vol.~12, p.~1473, 1979.

\bibitem{LegendreMiglio13}
D.~Legendre and M.~Maglio, ``Numerical simulation of spreading drops,'' {\em
  Colloids and Surfaces A: Physicochemical and Engineering Aspects}, vol.~432,
  pp.~29--37, 2013.

\bibitem{Courbin09}
L.~Courbin, J.~C. Bird, M.~Reyssat, and H.~A. Stone, ``Dynamics of wetting:
  from inertial spreading to viscous imbibition,'' {\em Journal of Physics:
  Condensed Matter}, vol.~21, p.~464127, 2009.

\bibitem{Bird08}
J.~C. Bird, S.~Mandre, and H.~A. Stone, ``Short-time dynamics of partial
  wetting,'' {\em Physical review letters}, vol.~100, p.~234501, 2008.

\bibitem{Carlson11}
A.~Carlson, M.~Do-Quang, and G.~Amberg, ``Dissipation in rapid dynamic
  wetting,'' {\em Journal of Fluid Mechanics}, vol.~682, pp.~213--240, 2011.

\bibitem{Winkels12}
K.~G. Winkels, J.~H. Weijs, A.~Eddi, and J.~H. Snoeijer, ``Initial spreading of
  low-viscosity drops on partially wetting surfaces,'' {\em Physical Review E},
  vol.~85, p.~055301, 2012.

\bibitem{JUWELS}
{J\"{u}lich Supercomputing Centre}, ``{JUWELS: Modular Tier-0/1 Supercomputer
  at the J\"{u}lich Supercomputing Centre},'' {\em Journal of large-scale
  research facilities}, vol.~5, no.~A135, 2019.

\end{thebibliography}

\end{document}